\documentclass[twocolumn]{aastex631} 
\graphicspath{{./}{figures/}}

\usepackage{amsmath}
\usepackage{bm}
\usepackage{color}
\usepackage{hyperref}
\makeatletter
\DeclareRobustCommand{\rvdots}{
  \vbox{
    \baselineskip4\p@\lineskiplimit\z@
    \kern-\p@
    \hbox{.}\hbox{.}\hbox{.}
  }}
\makeatother

\begin{document}
\newcommand{\dm}{\mathrm{DM}}
\newcommand{\snr}{\mathrm{S/N}}
\newcommand{\sfr}{\mathrm{SFR}}
\newcommand{\ks}{\textcolor{black}}
\newcommand{\kwm}{\textcolor{purple}}
\newcommand{\question}{\textcolor{red}}

\newcommand{\volrate}{$7.3^{+8.8}_{-3.8} \times 10^{4}$}
\newcommand{\volratenum}{$7.3^{+8.8}_{-3.8}$}
\newcommand{\gammaval}{$-1.3^{+0.7}_{-0.4}$}
\newcommand{\Echar}{$2.38^{+5.35}_{-1.64} \times 10^{41}$}
\newcommand{\n}{$0.96^{+0.81}_{-0.67}$}
\newcommand{\alphaval}{$-1.39^{+0.86}_{-1.19}$}
\newcommand{\muhost}{$1.93^{+0.26}_{-0.38}$}
\newcommand{\medhost}{$84^{+69}_{-49}$}
\newcommand{\sigmahost}{$0.41^{+0.21}_{-0.20}$}
\newcommand{\stddevhost}{$174^{+319}_{-128}$}

\newcommand{\nrateinterp}{$1.72^{+1.48}_{-1.10}$}
\newcommand{\alphavalrateinterp}{$-1.10^{+0.67}_{-0.99}$}

\title{Inferring the Energy and Distance Distributions of Fast Radio Bursts using the First CHIME/FRB Catalog}

\author[0000-0002-6823-2073]{Kaitlyn Shin}
  \affiliation{MIT Kavli Institute for Astrophysics and Space Research, Massachusetts Institute of Technology, 77 Massachusetts Ave, Cambridge, MA 02139, USA}
  \affiliation{Department of Physics, Massachusetts Institute of Technology, 77 Massachusetts Ave, Cambridge, MA 02139, USA}
\author[0000-0002-4279-6946]{Kiyoshi W.~Masui}
  \affiliation{MIT Kavli Institute for Astrophysics and Space Research, Massachusetts Institute of Technology, 77 Massachusetts Ave, Cambridge, MA 02139, USA}
  \affiliation{Department of Physics, Massachusetts Institute of Technology, 77 Massachusetts Ave, Cambridge, MA 02139, USA}
\author[0000-0002-3615-3514]{Mohit Bhardwaj}
  \affiliation{McGill Space Institute, McGill University, 3550 rue University, Montr\'eal, QC H3A 2A7, Canada}
  \affiliation{Department of Physics, McGill University, 3600 rue University, Montr\'eal, QC H3A 2T8, Canada}
\author[0000-0003-2047-5276]{Tomas Cassanelli}
  \affiliation{Department of Electrical Engineering, Universidad de Chile, Av. Tupper 2007, Santiago 8370451, Chile             }
\author[0000-0002-3426-7606]{Pragya Chawla}
  \affiliation{Anton Pannekoek Institute for Astronomy, University of Amsterdam, Science Park 904, 1098 XH Amsterdam, The Netherlands}
\author[0000-0001-7166-6422]{Matt Dobbs}
  \affiliation{Department of Physics, McGill University, 3600 rue University, Montr\'eal, QC H3A 2T8, Canada}
  \affiliation{McGill Space Institute, McGill University, 3550 rue University, Montr\'eal, QC H3A 2A7, Canada}
\author[0000-0003-4098-5222]{Fengqiu Adam Dong}
  \affiliation{Department of Physics and Astronomy, University of British Columbia, 6224 Agricultural Road, Vancouver, BC V6T 1Z1 Canada}
\author[0000-0001-8384-5049]{Emmanuel Fonseca}
  \affiliation{Department of Physics and Astronomy, West Virginia University, PO Box 6315, Morgantown, WV 26506, USA}
  \affiliation{Center for Gravitational Waves and Cosmology, West Virginia University, Chestnut Ridge Research Building, Morgantown, WV 26505, USA}
\author[0000-0002-3382-9558]{B.~M.~Gaensler}
  \affiliation{Dunlap Institute for Astronomy \& Astrophysics, University of Toronto, 50 St.~George Street, Toronto, ON M5S 3H4, Canada}
  \affiliation{David A.~Dunlap Department of Astronomy \& Astrophysics, University of Toronto, 50 St.~George Street, Toronto, ON M5S 3H4, Canada}
\author[0000-0002-3654-4662]{Antonio Herrera-Martín}
  \affiliation{David A.~Dunlap Department of Astronomy \& Astrophysics, University of Toronto, 50 St.~George Street, Toronto, ON M5S 3H4, Canada}
  \affiliation{Department of Statistical Sciences, University of Toronto, 700 University Ave., Toronto, ON M5G 1Z5, Canada}
\author[0000-0003-4810-7803]{Jane Kaczmarek}
  \affiliation{Dominion Radio Astrophysical Observatory, Herzberg Research Centre for Astronomy and Astrophysics, National Research Council Canada, PO Box 248, Penticton, BC V2A 6J9, Canada}
\author[0000-0001-9345-0307]{Victoria Kaspi}
  \affiliation{Department of Physics, McGill University, 3600 rue University, Montr\'eal, QC H3A 2T8, Canada}
  \affiliation{McGill Space Institute, McGill University, 3550 rue University, Montr\'eal, QC H3A 2A7, Canada}
\author[0000-0002-4209-7408]{Calvin Leung}
  \affiliation{MIT Kavli Institute for Astrophysics and Space Research, Massachusetts Institute of Technology, 77 Massachusetts Ave, Cambridge, MA 02139, USA}
  \affiliation{Department of Physics, Massachusetts Institute of Technology, 77 Massachusetts Ave, Cambridge, MA 02139, USA}
\author[0000-0003-2095-0380]{Marcus Merryfield}
  \affiliation{Department of Physics, McGill University, 3600 rue University, Montr\'eal, QC H3A 2T8, Canada}
  \affiliation{McGill Space Institute, McGill University, 3550 rue University, Montr\'eal, QC H3A 2A7, Canada}
\author[0000-0002-2551-7554]{Daniele Michilli}
  \affiliation{MIT Kavli Institute for Astrophysics and Space Research, Massachusetts Institute of Technology, 77 Massachusetts Ave, Cambridge, MA 02139, USA}
  \affiliation{Department of Physics, Massachusetts Institute of Technology, 77 Massachusetts Ave, Cambridge, MA 02139, USA}
\author[0000-0002-3777-7791]{Moritz M\"{u}nchmeyer}
  \affiliation{Department of Physics, University of Wisconsin-Madison, 1150 University Ave, Madison, WI 53706, USA}
\author[0000-0002-8912-0732]{Aaron B.~Pearlman}
  \affiliation{Department of Physics, McGill University, 3600 rue University, Montr\'eal, QC H3A 2T8, Canada}
  \affiliation{McGill Space Institute, McGill University, 3550 rue University, Montr\'eal, QC H3A 2A7, Canada}
\author[0000-0001-7694-6650]{Masoud Rafiei-Ravandi}
  \affiliation{Department of Physics, McGill University, 3600 rue University, Montr\'eal, QC H3A 2T8, Canada}
  \affiliation{McGill Space Institute, McGill University, 3550 rue University, Montr\'eal, QC H3A 2A7, Canada}
\author[0000-0002-2088-3125]{Kendrick Smith}
  \affiliation{Perimeter Institute of Theoretical Physics, 31 Caroline Street North, Waterloo, ON N2L 2Y5, Canada}
\author[0000-0001-9784-8670]{Ingrid Stairs}
  \affiliation{Department of Physics and Astronomy, University of British Columbia, 6224 Agricultural Road, Vancouver, BC V6T 1Z1 Canada}
\author[0000-0003-2548-2926]{Shriharsh P.~Tendulkar}
  \affiliation{Department of Astronomy and Astrophysics, Tata Institute of Fundamental Research, Mumbai, 400005, India}
  \affiliation{National Centre for Radio Astrophysics, Post Bag 3, Ganeshkhind, Pune, 411007, India}
\newcommand{\allacks}{
A.B.P. is a McGill Space Institute (MSI) Fellow and a Fonds de Recherche du Quebec - Nature et Technologies (FRQNT) postdoctoral fellow.
A.H.M is supported by the CANSSI CRT program.
B.M.G. is supported by an NSERC Discovery Grant (RGPIN-2022-03163), and by the Canada Research Chairs (CRC) program. 
C.L. was supported by the U.S. Department of Defense (DoD) through the National Defense Science \& Engineering Graduate Fellowship (NDSEG) Program.
F.A.D is supported by the U.B.C Four Year Fellowship.
FRB research at UBC is supported by an NSERC Discovery Grant and by the Canadian Institute for Advanced Research.
K.S. is supported by the NSF Graduate Research Fellowship Program.
K.W.M. is supported by NSF grants 2008031 and 2018490.
M.B. is supported by an FRQNT Doctoral Research Award.
M.D. is supported by a Canada Research Chair, NSERC Discovery Grant, CIFAR, and by the FRQNT Centre de Recherche en Astrophysique du Qu\'ebec (CRAQ).
M.Me. is supported by an NSERC PGS-D award.
MM is supported by DOE grant DE-SC0022342.
SPT is a CIFAR Azrieli Global Scholar in the Gravity and Extreme Universe Program.
V.M.K. holds the Lorne Trottier Chair in Astrophysics \& Cosmology, a Distinguished James McGill Professorship, and receives support from an NSERC Discovery grant (RGPIN 228738-13), from an R. Howard Webster Foundation Fellowship from CIFAR, and from the FRQNT CRAQ.
}


\correspondingauthor{Kaitlyn Shin}
\email{kshin@mit.edu}

\begin{abstract}

Fast radio bursts (FRBs) are brief, energetic, typically extragalactic flashes of radio emission whose progenitors are largely unknown.
Although studying the FRB population is essential for understanding how these astrophysical phenomena occur, such studies have been difficult to conduct without large numbers of FRBs and characterizable observational biases.
Using the recently released catalog of 536 FRBs published by the Canadian Hydrogen Intensity Mapping Experiment/Fast Radio Burst (CHIME/FRB) collaboration,
we present a study of the FRB population that also calibrates for selection effects.
Assuming a Schechter function, we infer a characteristic energy cut-off of $E_\mathrm{char} =$~\Echar~erg and a differential power-law index of $\gamma =$ \gammaval.
Simultaneously, we infer a volumetric rate of [\volratenum(stat.)$^{+2.0}_{-1.8}$(sys.)]$\times 10^4$~Gpc$^{-3}$~yr$^{-1}$ 
above a pivot energy of 10$^{39}$~erg and below a scattering timescale of 10 ms at 600~MHz,
and find we cannot significantly constrain the cosmic evolution of the FRB population with star formation rate.
Modeling the host dispersion measure (DM) contribution as a log-normal distribution and assuming a total Galactic contribution of 80 pc cm$^{-3}$, we find a median
value of $\dm_\mathrm{host} =$~\medhost~pc~cm$^{-3}$, comparable with values typically used in the literature.
Proposed models for FRB progenitors should be consistent with the energetics and abundances of the full FRB population predicted by our results.
Finally, we infer the redshift distribution of FRBs detected with CHIME, which will be tested
with the localizations and redshifts enabled by the upcoming CHIME/FRB Outriggers project.
\end{abstract}


\section{Introduction}
\label{sec:intro}

Ever since fast radio bursts (FRBs) were first discovered \citep{lorimer+2007,thornton+2013}, their mysterious origin has been an open question with no simple resolution.
Studying FRBs as a population can provide valuable insight into their nature, as well as their utility for cosmological applications 
\citep[e.g.,][]{mcquinn2014, zheng+2014_frbigm, zhou+2014, masuisigurdson2015,akahori+2016, madhavacheril+2019_ksz}.
A natural way to study FRBs is through inferring their luminosity function, which can tell us about observables such as volumetric rate and maximum FRB energy \citep{luo2020lumin}.
The inferred parameters can be compared with those of other known astrophysical events to better contextualize the question of FRB origins.
However, populations studies of FRBs are challenging --- as a population, FRBs have been observed in a variety of observational contexts; it is yet unknown how many distinct populations there may be, and whether there is a dominant progenitor channel \citep{james+2021_sfr}.
For example, some FRBs have been observed to repeat, while many more have only been observed as apparent one-off bursts.
Indeed, recent studies have put forth the possibility, on the basis of observational properties such as pulse widths and observed frequency bandwidths, that repeaters and non-repeaters may comprise two distinct populations of FRBs \citep[e.g.,][]{cui+2021, pleunis+2021_morphology}.

Furthermore, FRBs have been found to be associated with a variety of types of galaxies,
and in some cases, localized to Galactic regions with $\lesssim$arcsecond precision.
Repeaters have been shown to originate both within star-forming regions \citep[e.g.,][]{marcote+2020, nimmo+2022_R67_evnloc} and offset by $\sim$250~pc from the closest knot of local star formation \citep[e.g.,][]{tendulkar+2021_R3offset}.
By contrast, the repeating FRB 20200120E was localized to a globular cluster associated with M81 \citep{bhardwaj+2021_m81, kirsten+2022_m81}, surprising given that globular clusters host a much older stellar population than do star-forming regions.
FRB-like bursts have also been observed from the Galactic magnetar SGR~1935+2154 \citep{chimefrb_sgr1935, bochenek+2020_stare2_sgr1935, kirsten+2021_sgr1935}.
It is possible that more such FRB-like events occur in the Milky Way, but remain undetected due to interstellar medium (ISM) scattering \citep{goharflynn2022_MWfrbs}.
FRB localizations are not limited to repeating bursts;
apparent one-offs have also been localized to host galaxies, 
and in some cases, the localizations are precise enough to be associated with regions of low star formation rate as well \citep[e.g.,][]{bannister+2019_oneoff_loc,heintz+2020,bhandari+2020,li+zhang2020,mannings+2021}.
A comprehensive overview of the breadth of such localization results can be found in 
\citet{petroff+2022_frbreview}.

Nonetheless, there are many more FRBs that have been detected than just the $\sim$20 FRBs robustly associated with host galaxies.
By 2020, there were over 100 verified FRBs\footnote{\url{https://www.frbcat.org}} \citep{petroff+2016_frbcat},
and as of writing, the number of verified FRB events is $\sim$800; a full catalog of FRBs is maintained at the Transient Name Server (TNS).\footnote{\url{https://wis-tns.weizmann.ac.il}}
Based on these numbers alone, it is clear that studying FRBs as a population based only on the sub-sample of FRBs with precise localizations has strong limitations --- that is, optical follow-up is time-intensive and expensive, and cannot currently be feasibly pursued for the hundreds of detected FRBs.
There is also valuable information possessed by the many FRBs that do not have host galaxy associations, e.g., FRB energetics and abundances.
Using a larger sample of FRBs adds valuable information, and should be very beneficial for statistically constraining how FRB energies are distributed.
However, the central challenge to measuring the FRB luminosity function is, again, the lack of precise distance (localization) information for the majority of FRBs.
Without a good handle on distance, the FRB energetics, distances, and volumetric abundances are all degenerate. 


For FRBs, the dispersion measure ($\dm$) is the (imperfect) distance proxy that allows us to break this modeling degeneracy.
Since we expect $\dm$ to correlate with distance, while brightness will anti-correlate,
any correlation/anti-correlation in modeling between
$\dm$ and brightness
should be a distance effect, thus providing a distance scale.
As the uncertainties when modeling FRB energetics, distances, and abundances all affect each other, it is important to work with these properties simultaneously to minimize bias.
\citet{luo2020lumin} and \citet{james+2021_zDM} have recently implemented this approach to studying the FRB population.

Previously, in order to conduct large-number statistical studies of the FRB population, it was necessary to combine burst detections from multiple surveys or telescopes such as the Australian Square Kilometre Array Pathfinder (ASKAP) or the Parkes radio telescope \citep[e.g.,][]{luo2020lumin, james+2021_zDM}.
Using a heterogeneously observed sample of FRBs introduces differing, non-uniform selection effects that are difficult to calibrate.
This motivates the use of FRBs observed from a single observing campaign to study the FRB population as a whole.
Recently, the Canadian Hydrogen Intensity Mapping Experiment/Fast Radio Burst (CHIME/FRB) collaboration released a catalog of 536 FRBs, the largest sample of bursts detected thus far in a single survey, and hereafter referred to as ``Catalog~1'' \citep{catalog1}.
With Catalog~1, we now possess the ability to conduct a statistical study with uniform selection effects and probe the intrinsic FRB population with greater numbers and logistical simplicity.
Crucially, the selection effects are carefully characterizable with the use of the CHIME/FRB injections system \citep{merryfield+2022}.
With 536 bursts, the Catalog~1 sample dominates the currently available sample of observed FRBs.

In this paper, we present a study of FRB population parameters, including the intrinsic luminosity function, derived from the sample of Catalog~1 bursts.
We fit a joint brightness--$\dm$ model developed by \citet{james+2021_zDM}, and use formalism that accounts for selection effects detailed by \citet{catalog1}.
The outline of this paper is as follows.
In Section~\ref{sec:data}, we describe the datasets used for this analysis.
Section~\ref{sec:model} outlines key components of the model we are fitting to the Catalog~1 dataset, using the methodology described in Section~\ref{sec:methodology}.
Results are presented in Section~\ref{sec:results} and discussed in Section~\ref{sec:discussion}.
Implications of our results for the CHIME/FRB Outriggers project are presented in Section~\ref{sec:outriggers}, and we conclude in Section~\ref{sec:conclusion}.
We adopt Planck cosmological parameters \citep{planck2018vi} throughout this analysis.

\section{Dataset}
\label{sec:data}

The data used in this work consist of observed bursts from the CHIME/FRB Catalog~1 sample, corrected for selection effects in order for our analyses to accurately reflect properties of the intrinsic population of FRBs as much as possible.
The following subsections will detail the datasets used for this work.

\subsection{CHIME/FRB Catalog~1 observations}
\label{subsec:cat1data}

CHIME/FRB is a collaboration that uses the Canadian Hydrogen Intensity Mapping Experiment \citep[CHIME;][]{chime_sys_overview}, located at the Dominion Radio Astronomical Observatory (DRAO).
The telescope consists of four 20-m $\times$ 100-m cylindrical paraboloid reflectors,
each oriented N-S and populated with 256 equispaced dual-linear-polarization antennae.
This telescope is sensitive to bursts in the frequency range 400--800~MHz.
The CHIME/FRB instrument is designed such that it observes the entire sky at declination $>$ -11$^\circ$ as it transits above each day;
such a field of view is larger than other radio surveys or telescopes such as ASKAP or Parkes.
Thus, CHIME/FRB is uniquely equipped to observe a unparalleled number of FRBs.
Further technical details of the CHIME/FRB system can be found by \citet{chimefrb_sys_overview}.

The Catalog~1 sample released by CHIME/FRB contains 536 bursts observed between from 2018 July 25 to 2019 July 1.\footnote{This sample has been made publicly available at \url{https://chime-frb.ca/catalog}.}
In the work presented by \citet{catalog1}, the population analysis of Catalog~1 bursts considered six observed FRB properties measured for each burst:
fluence ($F$), dispersion measure ($\dm$), scattering timescale ($\tau$), pulse width ($w$), spectral index (denoted with $\gamma$ in \citet{catalog1}, but denoted with $\alpha$ in this paper), and spectral running ($r$).
These observed properties were modeled by a least-squares fitting routine called \texttt{fitburst}\footnote{The codebase for  is not yet open-source but based on the model formalism used by \citep{masui+2015}. A description of \texttt{fitburst} as developed and used by the CHIME/FRB collaboration will be presented by Fonseca et al. (in prep).}
which processes the intensity data of FRBs offline and fits the two-dimensional dynamic spectrum with a 2D analytic model.
For bursts with one or more components $i$, \texttt{fitburst} models $\dm$, burst time of arrival ($t_{\mathrm{arr}, i}$), signal amplitude ($A_i$),
temporal width ($w_i$), power-law spectral index ($\alpha_i$), spectral running ($r_i$), and scattering timescale ($\tau$).
The mathematical definitions of these components can be found in Section 3.3 of \citet{catalog1}.
The same $\dm$ and $\tau$ are assumed for all sub-bursts, and we also assume $\tau \propto \nu^{-4}$ \citep{lang1971, lorimerkramer2012_handbook}, with 600~MHz as the reference frequency for scattering.
For the small minority of bursts with complex structure in Catalog~1, if there was high enough $\snr$, \texttt{fitburst} was used to optimize the structure.
Dimmer bursts may have been over-dedispersed, especially for the bursts with downward-drifting structure, but the effect is likely minimal (i.e., no more than a few $\dm$ units).
For each FRB, \texttt{fitburst} was run twice: once with $\tau$ fixed to 0, and once allowing $\tau$ to vary. 
In order to choose the preferred model, an $F$-test\footnote{$F$-tests are good for comparing the statistics of nested models, and the model for bursts with scattering can be interpreted as nested from the model for bursts without scattering.} was used to compare the $\chi^2$ values from both models, adopting a threshold of $p < 0.001$ to determine the significance of scattering.
Electron density models of the Milky Way $\dm$, NE2001 \citep{ne2001} and YMW2016 \citep{ymw2016}, were used in combination with the $\dm$ value determined by \texttt{fitburst} to validate the extragalactic nature of each FRB.
Further details about \texttt{fitburst} can be found in Section~3.3 of \citet{catalog1}.

One further thing to note about CHIME/FRB observations is that fluence measurements are uncertain estimates of the true fluence, limited primarily by burst localization uncertainty and, to a lesser extent, by beam model uncertainty.
While these fluence measurements are biased low \citep[Section 3.4 in][]{catalog1}, our analysis is not affected by this systematic
because we use $\snr$ as a more reliable observable proxy to study fluence.
Further discussion of our treatment of fluence values is in Sections~\ref{subsec:fiducial_distrs} and \ref{subsec:compare_model_to_obs}.

\subsection{CHIME/FRB Catalog~1 injections}
\label{subsec:cat1injs}

In order to properly characterize and account for instrumental effects when 
inferring the intrinsic distributions of FRB properties,
a synthetic signal injection system was designed and implemented \citep{merryfield+2022}.
Synthetically generated pulses were injected into the software detection system in order to allow for the accounting of real-time detection effects such as the RFI environment.
These pulses will be hereafter referred to as ``injections.''

There are some important properties to note for the population of synthetic bursts injected into the real-time detection pipeline, which are briefly noted here but elaborated on by \citet{catalog1}.
Of the $N_\mathrm{inj} = 5 \times 10^6$ total injected FRBs assigned random locations on the sky, some locations were automatically discarded: locations below the horizon, where the band-averaged primary beam response was $< 10^{-2}$, and where the response was $< 10^{-3}$ for all 1024 synthesized beams.
The fraction of surviving sky locations is $f_{\mathrm sky} = 0.0277$.
This ``forward-modeling'' method, using the beam models to simulate $f_\mathrm{sky}$, was employed to account for the beam response of the telescope on the sky.
Each of these FRBs had properties drawn from initial probability density functions $P_{\mathrm{init}}(F)$, $P_{\mathrm{init}}(\dm)$, $P_{\mathrm{init}}(\tau)$, $P_{\mathrm{init}}(w)$,
and $P_{\mathrm{init}}(\alpha, r)$, designed to fully and densely sample phase space of observed properties in the catalog.
Of these $5 \times 10^6$ injected FRBs, a cut was applied to events with little chance of being detected, and 96,942 events were ultimately scheduled for injection.
Due to system errors that affected an effectively random subset of injections, 84,697 bursts were successfully injected with an injection efficiency $\epsilon_{\mathrm{inj}} = 0.874$ in 2020 August, with 39,638 events detected and assigned a signal-to-noise ratio ($\snr$) by the CHIME/FRB detection pipeline.
Full technical details on how the injections are detected by the real-time pipeline can be found in \citet{catalog1} and \citet{merryfield+2022}.

\subsection{Determining the fiducial property distributions}
\label{subsec:fiducial_distrs}

A primary goal of the initial Catalog~1 populations analysis was to fit a fiducial model to the intrinsic FRB property distributions after correcting for selection effects.
Such a model would be an imperfect, but still reasonable, match to the data.
However,
although the Catalog~1 sample makes up the first large sample of FRBs observed in a single survey with uniform selection effects,
a number of bursts are affected by certain selection effects that are unquantifiable (e.g., non-nominal telescope operation).
If we want to characterize the population while measuring and compensating for selection effects, we must only use the bursts with the most robustly known statistics.
Therefore, bursts affected by the following criteria were excluded from both the Catalog~1 sample as well as from the corresponding injections campaign sample:
\begin{enumerate}
    \item Events detected during pre-commissioning, during epochs of low sensitivity, or on days with software upgrades. These events were correspondingly excluded from the Catalog~1 survey duration time, $\Delta t$.
    \item Far-sidelobe events, for which the primary beam response is poorly understood.
    \item Events with a detected $\snr$ $<$ 12, to avoid human inspection error affecting the completeness of identifying low $\snr$ FRBs.
    \item Events with $\dm < 1.5 \; \mathrm{max(DM_{NE2001}, DM_{YMW16})}$, to avoid mistakenly classifying extragalactic bursts as Galactic or vice versa.
    \item Events with $\dm < 100$~pc~cm$^{-3}$.
    \item Highly scattered events ($\tau > 10$ms at 600~MHz), which are poorly observationally constrained due to a strong selection bias against them from the CHIME instrument, which would dominate uncertainties.
    \item Repeat bursts after the first-detected burst from any known repeating FRBs in the sample. 
\end{enumerate}

The first exclusion criterion (requiring $\snr$ $\geq$ 12) dominates the 271 total excluded bursts in the populations analysis conducted by \citet{catalog1}.
All of the above exclusion criteria are used in this analysis as well;
for more information about these cuts, refer to Sections 5 and 6.1 in \citet{catalog1}.
One further exclusion criterion than was used by \citet{catalog1} is used in this analysis.
In addition to excluding events with $\dm < 100$~pc~cm$^{-3}$,
in order to minimize the variance in our results involved with the Milky Way contribution to the $\dm$,
we \textit{also} exclude events with $\dm_\mathrm{NE2001} > 100$~pc~cm$^{-3}$.
This was done because in our model, we take a fixed value of $\overline{\dm}_\mathrm{MW}$ to be 80~pc~cm$^{-3}$; further discussion on this can be found in Section~\ref{sec:varyDML}.
It is worth noting that this extra exclusion criteria brings the total Catalog~1 sample to 225 bursts, down from the sample of 265 bursts used by \citet{catalog1}.
Thus, hereafter, when referring to the Catalog~1 sample used for populations studies in this analysis, we are referring to the 225 bursts that survived all of these cuts.

To elaborate on exclusion criterion \#7, simply excising all known repeating bursts would not work as there is no way to know if an apparent one-off burst will later turn out to be a repeater.
Choosing the highest $\snr$ burst will bias the sample to higher fluences, and choosing ``random'' bursts of the repeat bursts also introduce a subtle bias as the detection trigger threshold is set to be more sensitive to possible repeat bursts, i.e., bursts detected in the same directions and DMs as previously detected FRBs.
Thus, to minimize bias (especially with respect to trigger thresholds), we include only the first-detected burst from any known FRB source that otherwise passes other cuts (i.e., $\snr > 12$).
As detailed by \citet{pleunis+2021_morphology}, there are differences in burst widths and bandwidths for observed repeaters and apparent non-repeaters, raising the question of whether it is fair to assume identical detection biases for these bursts.
However, applying all the exclusion criteria except for the exclusion criterion for repeaters excludes 306 bursts instead of 311.
Thus, even though the selection biases may differ for repeaters and apparent non-repeaters, the difference is unlikely to strongly affect this analysis.
As the sample of observed repeaters grows, further investigation into the differing selection biases for repeaters and apparent non-repeaters will be crucial; further discussion is in Section~\ref{subsec:repeaters_v_nonrepeaters}.

Using the response of the real-time detection pipeline to the injected bursts, in \citet{catalog1} we were able to determine the selection functions for fluence, $\dm$, scattering, and pulse width.
Also, as detailed by \citet{catalog1}, the injections system does not have the ability to forward model fluence measurement processes in a robust manner, and measurements of the observed fluence are currently limited in their certainty.
Thus, the intrinsic fluence distribution was studied by using the detection $\snr$ as a proxy.
The key assumptions that go into using $\snr$ as a proxy for fluence are (1) detection $\snr$ is strongly correlated with intrinsic fluence, and (2) we can statistically accurately forward model how fluence maps to detection $\snr$ with the injections system.
As detailed in Section~\ref{subsec:compare_model_to_obs}, the same methodology is employed in this analysis.
Additionally, although spectral index and spectral running were two properties also assigned to each injected pulse and observed in each real burst, due to the correlated nature of those properties, a functional form was not fit to the distribution $P(\alpha, r)$.

Alongside the selection functions, the following selection-corrected fiducial distributions were determined: $P_\mathrm{fid}(F)$, $P_\mathrm{fid}(\dm)$, $P_\mathrm{fid}(\tau)$, and $P_\mathrm{fid}(w)$.
For each of these respective properties, these fiducial distributions provide a reasonable description to the Catalog~1 data.
For details on their derivation, refer to Section~6.1 and Appendix~C of \citet{catalog1}.
Here we note that to obtain these distributions, a number of simplifying assumptions were made in the Catalog~1 analysis, notably that the $F$--$\dm$ distribution factorizes into a power law in $F$ and a free function of $\dm$.
In this paper, we improve upon this $F$--$\dm$ relation and use a more realistic one following the model detailed by \citet{james+2021_zDM}.
Consistent with the framework we used to obtain the fiducial distributions in Catalog~1,
we hold the distributions of $\tau$ and $w$ to the obtained fiducial models $P_\mathrm{fid}(\tau)$ and $P_\mathrm{fid}(w)$ while fitting for a new joint distribution of $F$ and $\dm$ here.
For simplicity of notation, we define the parameter set $\mathbf{\Theta} = \{\tau, w\}$ such that $P(\mathbf \Theta) = P(\tau) P(w)$, and $\int d\mathbf \Theta P(\mathbf \Theta) = 1$.

\section{Fluence--DM distribution model}
\label{sec:model}

The model detailed by \citet{james+2021_zDM} contains components that detail how both fluence (i.e., brightness) and $\dm$ depend on redshift (i.e., distance).
Contained within this populations model is information about FRB energetics, how the FRB population may evolve over cosmic time,
as well as the distance distribution as inferred from $\dm$ observations.

We define the joint \textit{rate} distribution of $F$ and $\dm$ as
\begin{equation}
    \label{eqn:rate_fdm_distr}
	 R(F, \dm | \bm{\lambda}) = \int dz R(F, z | \bm{\lambda_1}) P(\dm | z, \bm{\lambda_2}),
\end{equation}
where the parameters $\bm{\lambda}$ for the fluence$\dm$ distribution considered will be detailed shortly. 
Inside the integrand, we can see two specific components: the joint rate distribution of fluence and redshift, and the probability distribution of $\dm$ given redshift.
Appendix~\ref{sec:modeldeets} details how to derive $R(F, \dm | \bm{\lambda})$,
closely following \citet{james+2021_zDM}.
The key components of the model are as follows.

We model the FRB energy distribution with a \citet{schechter1976} function
\begin{equation}
    \label{eqn:schechter}
	P(E) dE \propto \frac{1}{E_\mathrm{char}} \left(\frac{E}{E_\mathrm{char}}\right)^\gamma \exp{\left[-\frac{E}{E_\mathrm{char}}\right]} dE,
\end{equation}
where $E_\mathrm{char}$ is the characteristic exponential cutoff energy and $\gamma$ is the differential power-law index.
Such a form for the energy/luminosity distribution of FRBs has also been considered by previous works \citep[e.g.,][]{fialkov+2018, luo2018lumin, luo2020lumin, niu+2021_FDM}.
Note that we are assuming $P(E)$ does not evolve with redshift.

We also model the evolution of the FRB population by smoothly scaling the star formation rate (SFR) with a power-law index $n$,
\begin{equation}
    \label{eqn:sfr_evoln}
	\Phi(z) = \frac{\Phi_0}{1+z}
	\left(\frac{\sfr(z)}{\sfr(0)} \right)^n,
\end{equation}
where $\Phi(z)$ represents the rate of FRBs per comoving volume above a fixed pivot energy $E_\mathrm{pivot}$, and SFR$(z)$ (Equation~\ref{eqn:SFR(z)}) is from \citet{madaudickinson2014}.
A pivot energy is required for us to be able to quote a rate of FRBs;
we choose a pivot energy of $10^{39}$ erg under the assumption of spectral bandwidth at 1~GHz (see also Appendix~\ref{sec:R(F,z)}).
This pivot energy is above the
threshold of ruled out minimum FRB energies
(i.e., \citealt{james+2021_zDM} rules out {$E_\mathrm{min} > 10^{38.5}$} erg at 90\% C.L.).
The parameter $\Phi_0$ is the volumetric rate in units Gpc$^{-3}$ yr$^{-1}$ at $z=0$.
The population redshift evolution with the star formation history (SFH) is parameterized by $n$, where $n=0$ would imply no evolution with cosmic SFH
and $n=1$ would imply evolution that perfectly traces the cosmic SFH.
This model also allows for the possibility of FRB progenitors that evolve faster than the SFH of the universe (e.g., $n>1$).
Thus, the free parameters in $\bm{\lambda_1}$ are
$\Phi_0$, $E_\mathrm{char}$, $\gamma$, the spectral index $\alpha$, and $n$.
It is worth noting that $\alpha$ comes into the model in the conversion from $E$ to $F$ in Equation~\ref{eqn:E(F)}, such that this model assumes the intrinsic $\alpha$ is the same for every FRB.
The ``rate interpretation'' of $\alpha$, further discussed in Section~\ref{sec:rate_interp_alpha}, Appendix~\ref{sec:rate_alpha_supplementals}, and by \citet{james+2021_zDM}, takes the other extreme where every FRB has a unique spectral index.
As FRBs have been observed to have a wide range of spectral properties, 
the treatment of $\alpha$ is a systematic uncertainty that is model-dependent.

The $\dm$ distribution accounts for contributions from the disk and halo of the Milky Way, cosmological contributions through the intergalactic medium (IGM; refer to Equation~\ref{eqn:dmcosmic} in Section~\ref{sec:P(DM|z)}), and contributions from the host galaxy of the FRB.
The Milky Way contribution to the $\dm$ is fixed to be $\overline{\dm}_{\mathrm{MW}}$ = 80~pc~cm$^{-3}$ (discussed in Section~\ref{sec:varyDML}), and
the host $\dm$ contribution is modeled with a log-normal distribution
\begin{equation}\label{eqn:P_DMhost'}
    P(\dm_{\mathrm{host}}') =
    \frac{1}{\dm_{\mathrm{host}}'}
    \frac{1}{\sigma_{\mathrm{host}} \sqrt{2 \pi}}
    e^{
    -\frac{(\log \dm_{\mathrm{host}}' - \mu_{\mathrm{host}})^2}
    {2 \sigma^2_{\mathrm{host}}}
    },
\end{equation}
where the host $\dm$ contribution is parameterized by $\mu_{\mathrm{host}}$ and $\sigma_{\mathrm{host}}$
and corrected for redshift with $\dm_{\mathrm{host}} = \dm_{\mathrm{host}}' / (1+z) $.
(Note, the parameters as presented in the exponential of Equation~\ref{eqn:P_DMhost'} are in natural log space.)
One thing that is not considered in this model is the fact that the ionized gas content of galaxy and galaxy ensembles evolve over the cosmic star formation history.
Thus, the $\dm$ distribution would also evolve with redshift such that the observed $\dm$ distribution differs from the rest-frame one.
While a simple scaling relation could be applied to model the cosmic evolution of $\dm_\mathrm{host}$ \citep[e.g.,][]{luo2018lumin}, we choose not to in order to minimize the complexity of the model while keeping the physical parameters necessary to encapsulate the dependence of fluence and $\dm$ on distance.
Thus, the free parameters in $\bm{\lambda_2}$ are $\mu_{\mathrm{host}}$ and $\sigma_{\mathrm{host}}$.

Note, $\dm_\mathrm{host}'$ is the log-normal variate parameterized by $\mu_\mathrm{host}$ and $\sigma_\mathrm{host}$.
Thus, $\mu_{\mathrm{host}}$ and $\sigma_{\mathrm{host}}$ are the expected value and standard deviation of log($\dm_\mathrm{host}'$), not of $\dm_\mathrm{host}'$ itself.
The median and standard deviation of $\dm_\mathrm{host}'$ can be obtained with the parameters $\mu_{\mathrm{host}}$ and $\sigma_{\mathrm{host}}$ as follows:\footnote{\url{https://sites.google.com/site/probonto} \citep{probonto}}
\begin{eqnarray}
    \mathrm{Median} \; &&P(\dm_\mathrm{host}') = \exp({\mu_\mathrm{host}})
    \\ \notag
    \mathrm{Std. dev.} \; &&P(\dm_\mathrm{host}') =
    \\
    &&\sqrt{ (\exp(\sigma_\mathrm{host}^2)-1) \exp(2 \mu_\mathrm{host} + \sigma_\mathrm{host}^2) }.
\end{eqnarray}
Throughout this work, to avoid confusion, we consider the log-normal variate $\dm_\mathrm{host}'$ so as to keep all values in physical units of pc~cm$^{-3}$.
We also quote the median of $P(\dm_\mathrm{host}')$ rather than the mean,
as it is a log-normal distribution that we expect to be skewed by the existence of atypical, but plausible, sources with high host $\dm$ contributions,
and the median is more robust against outliers in a skewed distribution.

\subsection{Subtleties involved with CHIME/FRB fluence}
\label{subsec:compare_model_to_obs}

The model we aim to fit to the Catalog~1 data is a joint rate distribution of $F$ and $\dm$, $R(F, \dm | \bm \lambda)$.
However, as noted in Sections~\ref{subsec:cat1data} and \ref{subsec:fiducial_distrs}, observed fluence measurements are uncertain estimates of the true fluence.
While the synthetic bursts are assigned a ``true'' fluence value before injection, there is no good way to forward-model how their fluence values would be ``measured'' by the real-time detection pipeline.
Thus, directly comparing the ``true'' fluence values of the injected bursts with the ``observed'' fluence values of the Catalog~1 bursts would be subject to systematics not yet fully characterized.
The better alternative is to use $\snr$ as a proxy for fluence when fitting for $R(F, \dm | \bm \lambda)$, as fluence is strongly correlated with $\snr$ with known calibration factors, with the beam response also taken into account \citep{merryfield+2022}.

Thus, we wish to make a prediction for the \textit{observed} (as opposed to intrinsic) CHIME/FRB Catalog~1 DM--$\snr$ distribution using a given fluence--$\dm$ model.
The next section will detail how we fit for the $\snr$--$\dm$ distribution, and by extension, obtain the parameters $\bm{\lambda}$ in $R(F, \dm)$.

\section{Methodology}
\label{sec:methodology}

The observed CHIME/FRB Catalog~1 $\snr$--$\dm$ distribution is denoted with $n_{ij}$, where $i$ indexes the $\snr$ bin and $j$ indexes the $\dm$ bin.
We want to compare the observable $n_{ij}$ to our model prediction $\zeta_{ij}$, which depends on parameters
$\bm \lambda = \{ \Phi_0, E_\mathrm{char}, \gamma, \alpha, n, \mu_\mathrm{host}, \sigma_\mathrm{host} \}$.

The candidate population model $\zeta_{ij}$ is parameter-dependent --- and thus model-dependent --- because we construct weights to convert the original single injected population model to any other population model.
The alternative approach would have been to conduct an injections campaign for every candidate population model we wished to test, which is logistically infeasible (the full duration of the injections campaign used in this analysis was over a month).
Whereas $n_{ij}$ is an unweighted histogram of the observed Catalog~1 data in $\snr$ and $\dm$ bins, $\zeta_{ij}$ is a \textit{weighted} histogram of the synthetic data ``detected'' by the real-time pipeline.
Each synthetic burst $m$ in a given bin $n_{ij}$ has a weight $W_m(F, \dm, \mathbf{\Theta}, \bm \lambda)$ assigned to it, i.e.,
\begin{equation}
    \label{eqn:mu_ij}
    \zeta_{ij} = \sum_{m=1}^{n_{ij}} W_m(F, \dm, \mathbf{\Theta}, \bm \lambda).
\end{equation}
Note also that the units of $\zeta_{ij}$ is counts per $n_{ij}$ bin.
If we divide each bin in $\zeta_{ij}$ by the bin area ($\Delta \snr \; \Delta \dm$), then we would obtain an \textit{observed} rate of bursts per $\snr$ per $\dm$, e.g., $R_\mathrm{obs}(\snr, \dm | \lambda)$.
This quantity is the prediction for the \textit{observed} (as opposed to intrinsic) CHIME/FRB Catalog~1 DM--$\snr$ distribution.

We employ Markov chain Monte Carlo (MCMC) techniques to sample the likelihood while simultaneously fitting for all seven parameters in $R(F, \dm | \bm \lambda)$.
We use the logarithmic binned Poisson likelihood \citep{pdg2020}:
\begin{equation}
    \log \mathcal{L}(\bm{\lambda})
    \propto
    \sum_{ij} n_{ij} \log(\zeta_{ij}) - \zeta_{ij}.
    \label{eqn:posterior}
\end{equation}
The likelihood was evaluated with 15 logarithmically spaced bins in $\snr$, ranging from 12--200, and 20 logarithmically spaced bins in $\dm$, ranging from 100--3500~pc~cm$^{-3}$.

The evaluation of the different candidate population models $\zeta_{ij}$ is possible through the construction of weights $W_m(F, \dm, \mathbf{\Theta}, \bm \lambda)$. The formalism is as follows.
Each injected event $m$ has its own parameter-dependent weight $W_m(F, \dm, \mathbf{\Theta}, \bm \lambda)$,
which means each weight depends on the properties of each synthetic burst.
These weights are constructed to be the ratio of how many bursts were injected to how many occurred on the sky during the survey for a given set of model parameters $\bm \lambda$.
Thus, assuming the parameters $\bm \lambda$ are true, then for every injected burst with properties $F$ and $\dm$, we should have instead injected a number equal to $W_m(F, \dm, \mathbf{\Theta}, \bm \lambda)$ in order to exactly match the observed sky population.

To obtain the property-dependent weights $W_m(F, \dm, \mathbf{\Theta}, \bm \lambda)$,
for each synthetic burst $m$ with properties $F$ and $\dm$, using the model parameters $\mathbf{\Theta}$ and $\bm \lambda$,
we divide the full injected population of bursts on the sky by the initial injected population of bursts:
\begin{eqnarray}
    W(F, \dm, \mathbf{\Theta}, \bm \lambda) &=& \frac{R(F, \dm, \mathbf{\Theta} | \bm \lambda) \Delta t }{R_\mathrm{init}(F, \dm, \mathbf{\Theta})  } \\
    &=& \frac{\Delta t f_\mathrm{sky}}{N_\mathrm{inj} \epsilon_\mathrm{inj}}
    \frac{R(F, \dm, \mathbf{\Theta} | \bm \lambda)}{P_\mathrm{init}(F, \dm, \mathbf{\Theta})}.
\end{eqnarray}
The quantity $R_\mathrm{init}(F, \dm, \mathbf{\Theta})$ is the full initial population of injected bursts, with units of bursts per $F$ per $\dm$ per injections campaign.
It is equivalent to the normalized initial population distribution of the injected synthetic bursts, $P_\mathrm{init} (F, \dm, \mathbf{\Theta})$,
multiplied by the total number of FRBs injected ($N_\mathrm{inj}$) corrected for the efficiency of injections ($\epsilon_{\mathrm{inj}}$) and the sky fraction ($f_{\mathrm{sky}}$).
That is,
\begin{equation}
    R_\mathrm{init}(F, \dm, \mathbf{\Theta})
    = \frac{N_\mathrm{inj} \epsilon_\mathrm{inj}}{f_\mathrm{sky}} P_\mathrm{init}(F, \dm, \mathbf{\Theta}).
\end{equation}
The model $R(F, \dm, \mathbf{\Theta} | \bm \lambda)$ gives a rate in bursts per year.
When multiplied by $\Delta t$, the survey duration in days ($\Delta t = 214.8$ days),
we can then get a rate of observed bursts per survey, which can be directly compared against $R_\mathrm{init}(F, \dm, \mathbf{\Theta})$.

Also, as mentioned at the end of Section~\ref{subsec:fiducial_distrs}, the distributions of $\mathbf \Theta$ were held to the fiducial models found in Catalog~1.
Therefore,
\begin{eqnarray}
    W(F, \dm, \mathbf{\Theta}, \bm \lambda) &=& \frac{\Delta t f_\mathrm{sky}}{N_\mathrm{inj} \epsilon_\mathrm{inj}}
    \frac{R(F, \dm | \bm \lambda)}{P_\mathrm{init}(F, \dm)}
    \frac{P_\mathrm{fid}(\mathbf \Theta)} {P_\mathrm{init}(\mathbf \Theta)},
\end{eqnarray}
where the fiducial distributions and initial distributions are all factorizable.

Each synthetic burst can have an associated weight because each burst has ``intrinsic'' $F$ and $\dm$ information associated with it.
Because each burst also has ``observed'' $\snr$ information, each weighted burst can be assigned a bin in $\zeta_{ij}$, thus allowing for $\zeta_{ij}$ to be model-dependent (i.e., dependent on $\bm \lambda$) and directly comparable to $n_{ij}$ in the likelihood function (Equation~\ref{eqn:posterior}).

Note that the equations in this section are simply a more explicit definition of the weights also defined in \citet{catalog1}, wherein
the weights were used only for obtaining the model parameters for $P_\mathrm{fid}(F)$, $P_\mathrm{fid}(\dm)$, and $P_\mathrm{fid}(\mathbf{\Theta})$.
For these distributions, the pre-factors used to accurately determine the rate ($\Delta t f_\mathrm{sky} / N_\mathrm{inj} \epsilon_\mathrm{inj}$) do not matter.

\section{Results}
\label{sec:results}

\begin{figure*}[htbp]
    \centering
    \includegraphics[width=\textwidth]{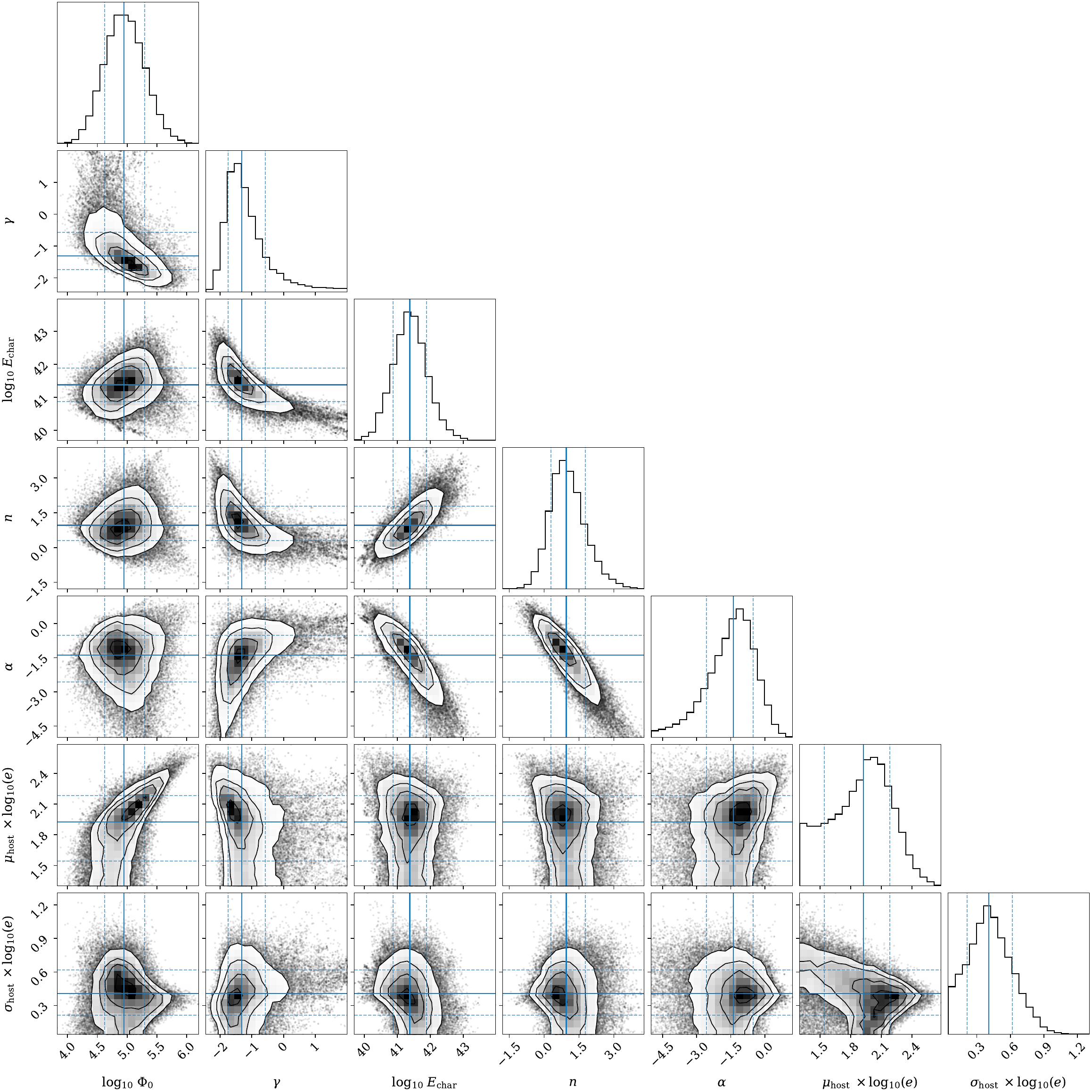}
    \caption{
        Corner plot of the results of the MCMC run, thinned by a factor of 20 for visualization purposes.
        Overlaid solid blue lines denote the median of the posterior distributions, with the dashed blue lines enclosing the central 68\% of the samples.
    }
    \label{fig:mcmc_corner}
\end{figure*}
Using the methodology described by \citet{catalog1} and elaborated on in Section~\ref{sec:methodology}, we use the injections system to calibrate out selection effects while fitting this full population model, $R(F,\dm | \bm \lambda)$, to the Catalog~1 data.
We use the \texttt{emcee} package \citep{dfm+2013_emcee} to generate MCMC samples of the posterior, using \texttt{scipy.optimize.curve\_fit} to obtain the initial parameters for the MCMC chains.
In order to ensure the posterior distribution is proportional to the likelihood in Equation~\ref{eqn:posterior},
all of our parameters are drawn from uniform priors.
We then run the MCMC sampler with 40 walkers, 2000 burn-in steps, and then 25000 steps after that.
While there is no simple way to ensure convergence, the first and second halves of the chains show similar posterior distribution morphology to each other, heuristically indicating confidence our chains have converged \citep{hogg_dfm2018_usingMCMC}.
Additionally, the chains are longer than 75 times the integrated autocorrelation time for all parameters,
with an acceptance fraction of $\approx$0.26 --- within the range of recommended values \citep{dfm+2013_emcee}.
The results can be seen in Figure~\ref{fig:mcmc_corner}, generated using the \texttt{corner} package \citep{dfm2016_corner}.\footnote{Data products are also made available at \url{https://github.com/kaitshin/CHIMEFRB-Cat1-Energy-Dist-Distrs}.}
We quote the median of the posterior distributions, along with the 16\% quantile as the lower 1$\sigma$ error bar and the 84\% quantile as the upper 1$\sigma$ error bar, as generally recommended by \citet{hogg_dfm2018_usingMCMC}.
These values are tabulated in Table~\ref{tab:param_results}.

In order to verify the robustness of the MCMC sampling, we also redo our analysis on ``mock'' data.
In order to get a mock CHIME/FRB Catalog~1 dataset,
we took the full initial population of injected FRBs and resampled them with weights according to the best-fit model parameter values.
We then ran the MCMC sampler with this mock Catalog~1 data and recovered all the best-fit model parameters to within 2$\sigma$ (most parameters to within 1$\sigma$.

\begin{deluxetable*}{ccc}
    \tablecaption{
    Table of results of the parameters fitted in $R(F, \dm)$
    (see Equation~\ref{eqn:rate_fdm_distr}).
    Best-fit results quote the median values of the posterior distributions, with the error bars containing the central 68\% of the samples.
    Where illustrative, fits are paired with corresponding derived physically meaningful quantities.
    }
    \label{tab:param_results}
    \tablehead{\colhead{Parameter} & \colhead{Uniform prior range} & \colhead{Best-fit result}} 
    
    \startdata
$\log_{10} \; \Phi_0 \: ^a$ & [$-0.96$, $6.43$] & $4.86^{+0.34}_{-0.33}$ \\
$\Phi_0$ & $\ldots$ & $7.3^{+8.8}_{-3.8} \times 10^{4}$ Gpc$^{-3}$ yr$^{-1}$ \\[0.5em]
$\gamma$ & [$-2.50$, $2.00$] & $-1.3^{+0.7}_{-0.4}$ \\[0.5em]
$\log_{10} \; E_{\mathrm{char}} \: ^a$ & [$38.00$, $49.00$] & $41.38^{+0.51}_{-0.50}$ \\
$E_{\mathrm{char}}$ & $\ldots$ & $2.38^{+5.35}_{-1.64} \times 10^{41}$ erg \\[0.5em]
$n$ & [$-2.00$, $8.00$] & $0.96^{+0.81}_{-0.67}$ \\[0.5em]
$\alpha$ & [$-5.00$, $5.00$] & $-1.39^{+0.86}_{-1.19}$ \\[0.5em]
$\mu_{\mathrm{host}} \; \times \log_{10}(e) \: ^a$ & [$1.30$, $2.70$] & $1.93^{+0.26}_{-0.38}$ \\
$\mathrm{Median} \; P(\mathrm{DM}\sp{\prime})$ & $\ldots$ & $84^{+69}_{-49} $ pc cm$^{-3}$ \\[0.5em]
$\sigma_{\mathrm{host}} \; \times \log_{10}(e) \: ^a$ & [$0.04$, $1.74$] & $0.41^{+0.21}_{-0.20}$ \\
$\mathrm{Std. dev.} \; P(\mathrm{DM}\sp{\prime})$ & $\ldots$ & $174^{+319}_{-128}$ pc cm$^{-3}$ \\
    \enddata
    \tablenotetext{a}{These parameters were fitted in natural $\log$ space but are presented in $\log_{10}$ space for ease of interpretation.}
\end{deluxetable*}

\begin{figure*}[htbp]
    \centering
    \includegraphics[width=\textwidth]{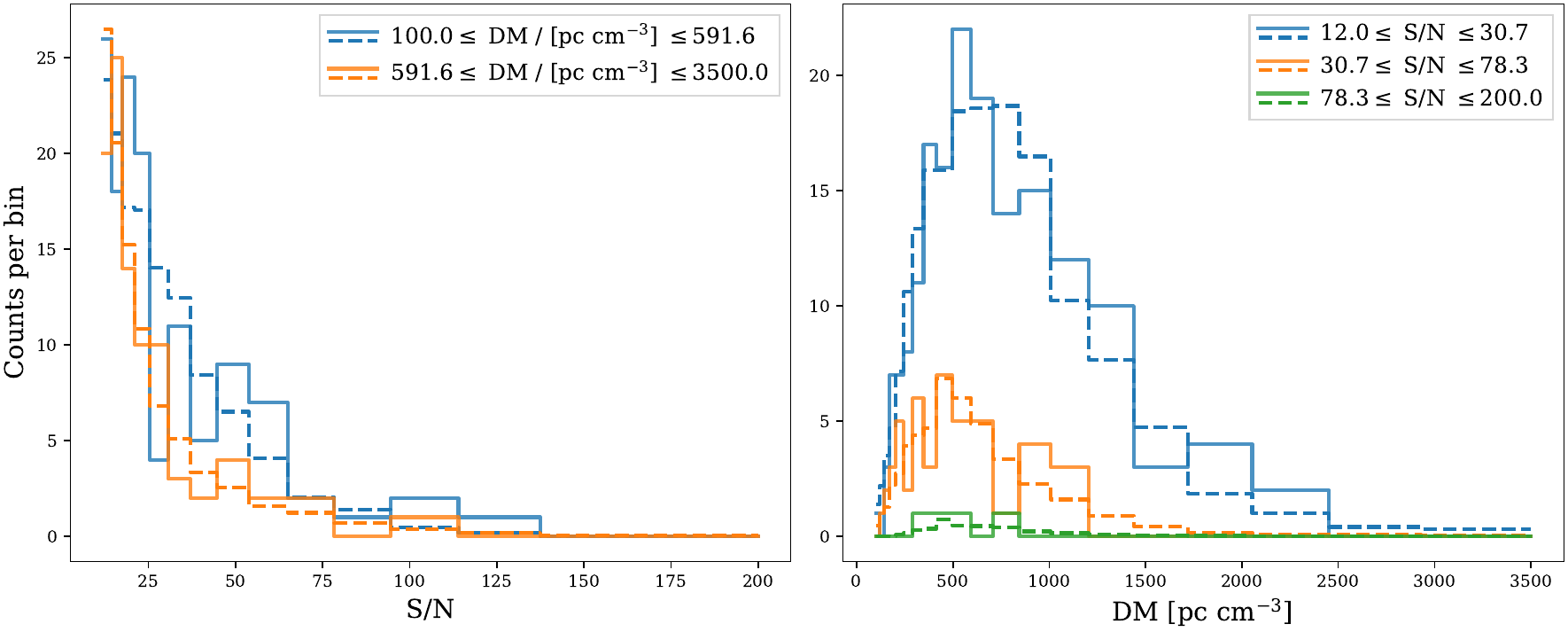}
    \caption{
        Observed distributions, denoted with solid lines, compared with best-fit model predictions, denoted with dashed lines.
        Each sample range is determined by the logarithmic binning described in Section~\ref{sec:methodology},
        and not by the number of events that fall in each bin.
        \textbf{Left panel:} The $\snr$ distribution is shown for a low-DM sample and a high-DM sample. Each shown $\dm$ range spans ten bins.
        \textbf{Right panel:} The $\dm$ distribution is shown in three $\snr$ samples. Each shown $\snr$ range spans five bins.
        The best-fit model predictions use the median of the posterior distributions of each parameter after the MCMC run.
        In every viewed slice, the model appears consistent with the observed data.
    }
    \label{fig:distr_pred_slices}
\end{figure*}

Figure~\ref{fig:distr_pred_slices} shows slices through the $\snr$--$\dm$ distribution of the observed Catalog~1 data $n_{ij}$ compared against the best-fit model prediction from the MCMC run, $\zeta_{ij}$, in each parameter space.
In both the low and high $\dm$ slices, the observed $\snr$ distribution peaks at lower values closer to the $\snr$ cutoff threshold, with negligibly few sources at the high-end of the $\snr$ distribution.
Additionally, when viewing the data in $\snr$ slices, there are more sources in the lower $\snr$ ranges than in the higher ones,
with the $\dm$ distribution appearing to peak at $\sim$500 pc cm$^{-3}$ for the slice most representative of the Catalog~1 data (the lowest $\snr$ sample).
Note the rarity of detecting exceptionally high $\snr$ events at any $\dm$ range.
In each view of the slices through the data, the best-fit model appears to be consistent with the observed data.

\begin{figure*}[htbp]
    \centering
    \includegraphics[width=\textwidth]{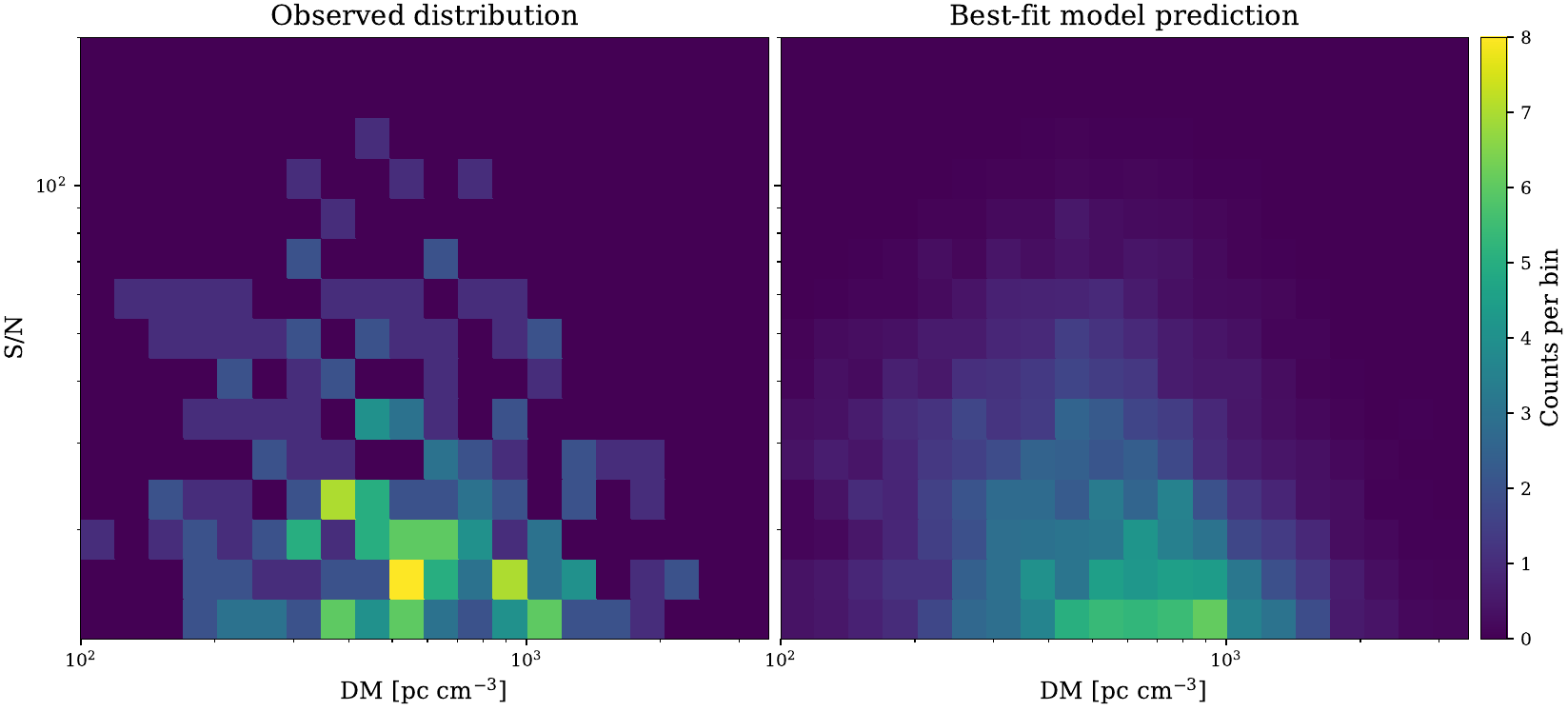}
    \caption{
        Comparison of the $\snr$--DM distribution between the observed data (left) and the best-fit model prediction (right).
        The best-fit model predictions use the median of the posterior distributions of each parameter after the MCMC run.
        The two appear to behave qualitatively similarly, in-line with expectations from Figure~\ref{fig:distr_pred_slices}.
    }
    \label{fig:distr_pred_hist}
\end{figure*}

In Figure~\ref{fig:distr_pred_hist}, one can see a side-by-side comparison of the 2D view of the observed $\snr$--$\dm$ distribution and model prediction.
Qualitatively, it is clear that the data appear to be consistent with the model and added Poissonian noise.
Both the observed data and the model prediction show that the majority of FRBs are detected at lower $\snr$ ranges, peaking around $\dm$ values of hundreds of~pc~cm$^{-3}$.

\subsection{Goodness of fit}
\label{subsec:goodnessoffit}

One advantage of using maximum likelihood methods with binned data, as was mentioned in Section~\ref{sec:methodology}, is that a goodness-of-fit statistic can be simultaneously obtained \citep{pdg2020}.
Maximizing the likelihood $\mathcal{L}(\bm{\lambda})$ is equivalent to maximizing the likelihood ratio $\xi (\bm{\lambda})$,
and thus is also equivalent to minimizing the quantity $-2 \log \xi (\bm{\lambda})$.
We define the quantity $T = -2 \log \xi (\bm{\lambda})$.
For Poissonian data, this expression takes the form
\begin{equation}
    \label{eqn:T}
    T = 2 \sum_{ij} 
    \left[ \zeta_{ij} - n_{ij} 
    + n_{ij} \log \left( \frac{n_{ij}}{\zeta_{ij}} \right)  \right].
\end{equation}

\begin{figure}[htbp]
    \centering
    \includegraphics[width=\columnwidth]{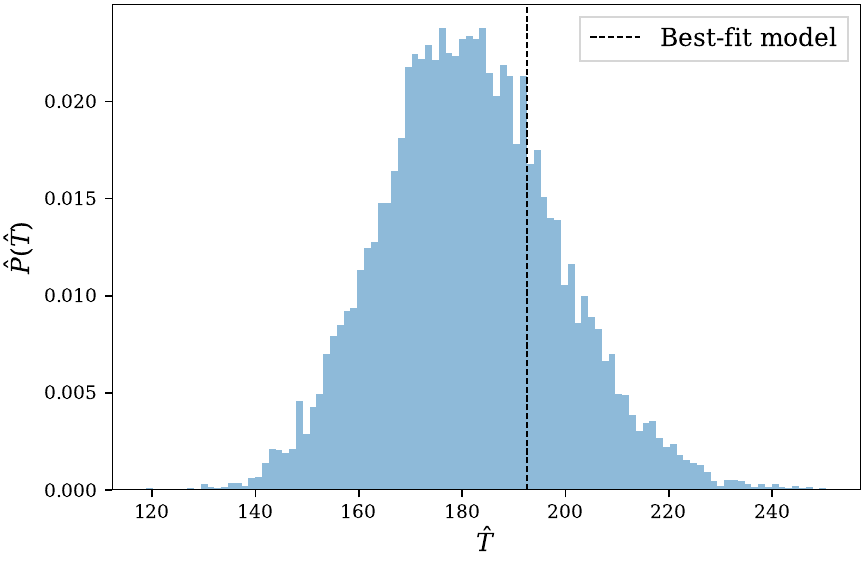}
    \caption{
    The estimated probability distribution $\hat P(\hat T)$ of the test statistic $T$, defined in Equation~\ref{eqn:T}, from Monte Carlo sampling.
    The value of this test statistic computed for our best-fit model prediction is overlaid in a vertical dashed line.
    Approximately 25\% of $\hat P(\hat T)$ lies to the right of the dashed line, giving an estimated $p$-value of 0.25.
    }
    \label{fig:mock_Ts}
\end{figure}

The smaller the value of $T$, the better the data $n_{ij}$ agree with the model $\zeta_{ij}$.
If Wilks' theorem holds, then the minimum of $T$ follows a $\chi^2$ distribution \citep{wilks1938}, and the value can be interpreted as a $p$-value.
However, one of the conditions of Wilks' theorem is that the sample size must be large, which our binned data do not satisfy.
Thus, to quantify the goodness of fit, we aim to estimate the null distribution of $T$ by Monte Carlo simulations.
We generate 10,000 Monte Carlo samples of Poissonian-distributed $n_{ij}$ from our $\zeta_{ij}$ and calculate the test statistic $T$.
The results of the Monte Carlo simulations are in Figure~\ref{fig:mock_Ts}.
Based on this statistic, over 100 runs, $\approx$25\% of the simulations have a worse fit than our best-fit model prediction.
In other words, we have estimated a $p$-value of 0.25 where in 10,000 different realizations of the data, the model $R(F,\dm | \bm \lambda)$ was a worse fit.
To reiterate, this $p$-value represents the fraction of data realizations drawn from the best-fit model that are a better fit to the data.
Therefore, a $p$-value of 0.25 indicates that the model is neither an anomalously bad nor good fit (i.e., overfit) to the data.
Thus, this estimated $p$-value is consistent with the conclusion that our model matches the data reasonably well, albeit with room for improvement in some combination of the model details, the collected data, and our treatment of observed selection effects.


\subsection{$R(F,z)$ parameters}

The parameters in $\bm{\lambda_1}$ are $\Phi_0$, $E_\mathrm{char}$, $\gamma$, $\alpha$, and $n$.
Contained within this component of the fit model are predictions about the energetics and abundances of FRBs based on the CHIME/FRB Catalog~1 sample.

We find a volumetric rate at $z=0$ above $E_{\mathrm{pivot}} = 10^{39}$~erg,
with our statistical uncertainties given by the MCMC. 
In the work presented by \citet{catalog1}, the net systematic uncertainty for the sky rate was~$+27\%$/$-25\%$; 
we assume the systematics propagate forward in a similar manner for the volumetric rate derived here. 
Thus, the volumetric rate is $\Phi_0 =$~[\volratenum(stat.)$^{+2.0}_{-1.8}$(sys.)] $\times 10^4$~Gpc$^{-3}$~yr$^{-1}$.
In \citet{catalog1}, the systematic errors are by far the largest for factors that affect the overall rate.
Given that in this analysis, the systematic errors are subdominant to the statistical errors for the rate calculation,
we assume that the remainder of the parameters are dominated by statistical errors as well.
Thus, for all other parameters we quote only statistical errors on our values.

The energy distribution within $R(F, z)$ is modeled with a Schechter function (Equation~\ref{eqn:schechter}), and we find best-fitting values of $E_{\mathrm{char}} =$~\Echar~erg and $\gamma =$~\gammaval,
where $\gamma$ is the differential power-law index.

Our best-fitting value for the spectral index in $F \propto \nu^\alpha$ is $\alpha =$ \alphaval.
However, one can also interpret $\alpha$ as a frequency-dependent rate;
see Section~\ref{sec:rate_interp_alpha} for details of this interpretation. 
For the parameter $n$ which parameterizes evolution with the cosmic SFH, we find a best-fitting value of $n =$~\n.
As can be seen in Figure~\ref{fig:mcmc_corner}, the parameters $\alpha$ and $n$ are strongly anti-correlated --- intuitively, the more strongly FRBs appear to evolve with (or faster than) the cosmic SFH (higher $n$), the more sources would originate from further away and appear fainter, thus leading to a steeper observed ``true spectral index.''
This degeneracy is stronger under the ``rate interpretation'' of $\alpha$.
The parameter for the cutoff energy $E_\mathrm{char}$ is also correlated (anti-correlated) with $n$ ($\alpha$), as the more energetic we expect FRBs can be, the more FRBs we should be able to observe further away (fainter).

\subsection{$P(\dm|z)$ parameters}

The parameters in $\bm{\lambda_2}$ are $\mu_{\mathrm{host}}$ and $\sigma_{\mathrm{host}}$.
Modeling the host galaxy contribution of the FRB to its $\dm$ as a log-normal distribution,
our best-fit values are $\mu_\mathrm{host} \times \log_{10}(e) =$ \muhost\ and $\sigma_\mathrm{host} \times \log_{10}(e) =$ \sigmahost.
These values correspond to a median $\dm_\mathrm{host}'$ value of \medhost~pc~cm$^{-3}$, with a standard deviation of \stddevhost~pc~cm$^{-3}$.

As can be seen in Figure~\ref{fig:mcmc_corner}, 
the allowed parameter space for $\mu_\mathrm{host}$ and the volumetric rate $\Phi_0$ has a complex morphology.
To intuitively explain some of this morphology, $\mu_\mathrm{host}$ sets the overall distance scale for FRBs -- the higher $\mu_\mathrm{host}$ is, the more we expect the $\dm$ contribution to be from the host galaxy rather than the IGM, and thus FRBs would be closer, less energetic, and more volumetrically abundant.

\subsubsection{Varying assumed $\overline{\dm}_{\mathrm{MW}}$}
\label{sec:varyDML}

When fitting for $P(\dm|z)$, we adopted $\overline{\dm}_{\mathrm{MW}}$~=~80~pc~cm$^{-3}$ as our fiducial value.
However, the Galactic contribution to the $\dm$ of an FRB has non-negligible uncertainties, especially when considering the Galactic halo.
Although we currently aim to minimize the uncertainty associated with the halo contribution by excluding sources with $\dm_\mathrm{NE2001}>100$~pc~cm$^{-3}$ (Section~\ref{subsec:fiducial_distrs}),
a more sophisticated analysis would take into account updated models of the Milky Way halo $\dm$ \citep[e.g.,][]{cook+2023_halo}.
Nonetheless, we wish to test how sensitive our analysis may be to different values of $\overline{\dm}_{\mathrm{MW}}$~=~80~pc~cm$^{-3}$.
Thus, we performed multiple more iterations of the MCMC analysis, varying $\overline{\dm}_{\mathrm{MW}}$ to between be 50~pc~cm$^{-3}$ and 100~pc~cm$^{-3}$ in increments of 10~pc~cm$^{-3}$.
We do these discrete iterations because otherwise, $\overline{\dm}_{\mathrm{MW}}$ as a free parameter would be degenerate with $\mu_\mathrm{host}$ \citep[see, e.g.,][]{macquart+2020_dmz}.
In fact, modifying the formalism to include $\overline{\dm}_{\mathrm{MW}}$ as a free parameter would likely have constraining power only after the sample of FRBs grows to reduce the already large statistical uncertainties already associated with $\mu_\mathrm{host}$ and $\sigma_\mathrm{host}$.
Having $\dm_{\mathrm{MW}}$ as a direction-dependent free parameter, and also having it be independent of $\mu_\mathrm{host}$, would require significant changes to the model formalism, and is beyond the scope of this analysis.

\begin{figure*}[htbp]
    \centering
    \includegraphics[width=\textwidth]{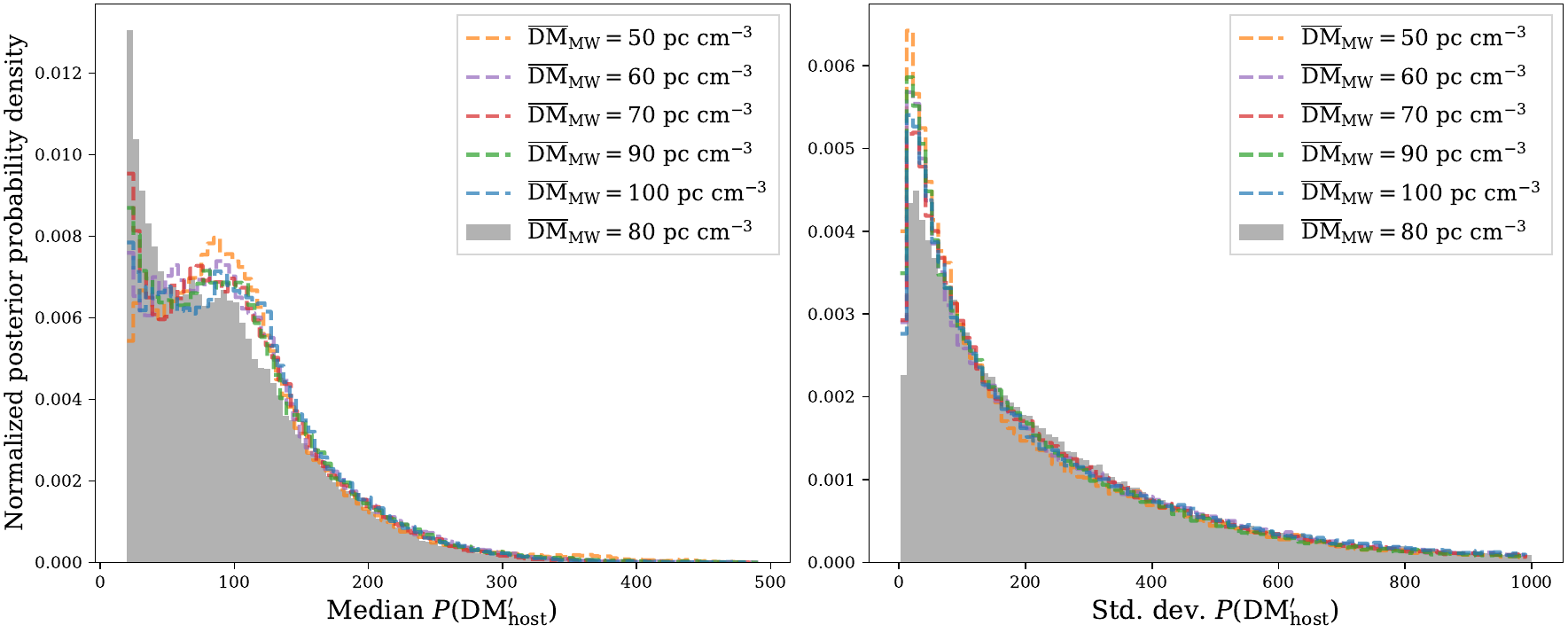}
    \caption{
        Overlaid posterior distributions from the MCMC runs run with 
        $\overline{\dm}_{\mathrm{MW}}$ = 50 pc cm$^{-3}$ (orange dotted-dashed lines), 80 pc cm$^{-3}$ (gray shaded region), and 100 pc cm$^{-3}$ (blue dashed lines).
        On the left is the normalized posterior distribution for $\mu_{\mathrm{host}}$
        and on the right is the normalized posterior distribution for $\sigma_{\mathrm{host}}$.
        When $\overline{\dm}_{\mathrm{MW}}$ is varied away from 80~pc~cm$^{-3}$, the resulting posterior distributions of $\mu_{\mathrm{host}}$ and of $\sigma_{\mathrm{host}}$
        appear insignificantly changed.
        Thus the parameters we find assuming a log-normal distribution for $\dm_\mathrm{host}$ are relatively robust to reasonable assumptions of $\overline{\dm}_{\mathrm{MW}}$.
    }
    \label{fig:DM_L_vary}
\end{figure*}

Predictably, the varying of this $\overline{\dm}_{\mathrm{MW}}$ assumption most affects the posterior distributions of the parameters that go into the log-normal distribution model of DM$_\mathrm{host}$, e.g.,
$\bm{\lambda_2}$ = \{$\mu_{\mathrm{host}}$, $\sigma_{\mathrm{host}}$\}.
None of the five other posterior distributions from the MCMC run appreciably changed.
Shown in Figure~\ref{fig:DM_L_vary} are the posterior distributions for the derived physically meaningful quantities corresponding to $\mu_{\mathrm{host}}$ and $\sigma_{\mathrm{host}}$ --- i.e., the median and the standard distribution of $P(\dm ^\prime_\mathrm{host})$, which is parameterized by $\mu_{\mathrm{host}}$ and $\sigma_{\mathrm{host}}$ --- from the runs where $\overline{\dm}_{\mathrm{MW}}$ was varied.
Comparing Figure~\ref{fig:DM_L_vary} to the results presented in Table~\ref{tab:param_results}, the 
variation of where the posterior distributions peak appears to be within 1$\sigma$ statistical uncertainties.
Thus, while we know our assumption of
$\overline{\dm}_{\mathrm{MW}}$ is imperfect, its effect on the modeling of $\dm_{\mathrm{host}}$ is sub-dominant to statistical uncertainties,
and it also does not appreciably affect the other fit parameters in $R(F, \dm | \bm \lambda)$.
As such, we continue with our fiducial value of $\overline{\dm}_{\mathrm{MW}}$~=~80~pc~cm$^{-3}$.

\subsection{``Rate interpretation'' of $\alpha$}
\label{sec:rate_interp_alpha}

Throughout this work, we have treated $\alpha$ as a \textit{true} FRB spectral index,
i.e., a spectral index for an individual FRB (observed as a broad-band burst) that also applies to the full population of studied FRBs.
The flux behaves as $F \sim \nu^\alpha$.
However, as \citet{james+2021_zDM} noted, $\alpha$ can also be interpreted as a frequency-dependent rate,
particularly when considering experiments with a similar bandwidth to that of the detected FRB.
As FRBs can be band-limited \citep[e.g.,][]{law+2017_frb121102, pleunis+2021_morphology},
this means that there are either a larger number of low-frequency FRBs, or that FRBs are stronger at low frequencies.
Under such an interpretation, the rate now behaves as $\Phi \sim \Phi(z) \nu^\alpha$, and through the changed $k$-correction, the factor of $(1+z)^\alpha$ is directly added to the source evolution (Equation~\ref{eqn:Phi(z)}), thus most directly affecting the parameter $n$.
It is likely that the most realistic treatment of $\alpha$ is more complicated than either single interpretation --- for example, it is reasonable to think FRBs may be both broad-band (as per the ``true spectral index'' interpretation) and have unique spectral indices (as per the ``rate interpretation'').
Thus the treatment of $\alpha$ remains a model-dependent systematic uncertainty.

For completeness, we redo the analysis, this time treating $\alpha$ as a frequency-dependent rate.
The complete results of the parameters can be found in 
Figure~\ref{fig:mcmc_corner_alpharate} and
Table~\ref{tab:param_results_alpharate}.
Of the parameters, $\alpha$ and $n$ had the most significant changes;
they become far more correlated under the ``rate interpretation'' of $\alpha$.
Under the ``true spectral index interpretation'' of $\alpha$, we obtain $\alpha =$~\alphaval\ and $n =$~\n;
under the ``rate interpretation'' of $\alpha$, we obtain $\alpha\sp{\prime} =$~\alphavalrateinterp\ and $n\sp{\prime} =$~\nrateinterp.

\section{Discussion}
\label{sec:discussion}

Using the population model developed by \citet{james+2021_zDM}, this analysis
simultaneously constrains the redshift, energy, and host $\dm$ distributions of FRBs using CHIME/FRB Catalog~1 data.
This is the first time such a population study has been conducted on a large sample of bursts from the same uniform survey while explicitly accounting for selection effects; previous works have had to combine bursts from different samples, while attempting to account for all the heterogeneous selection effects involved \citep[e.g.,][]{luo2020lumin, james+2021_zDM}.
Furthermore, it is the injections system of the CHIME/FRB instrument that crucially allows for the careful calibration of selection biases in a comprehensive end-to-end way that no other radio survey is currently able to conduct.
There are a total of seven free parameters fitted in this population model to the CHIME/FRB data; when interpreting $\alpha$ as a frequency-dependent rate, another set of fit parameters can be obtained.
In this section, we discuss our results and the property distributions they describe.
Notably, in Section~\ref{subsec:repeaters_v_nonrepeaters}, we discuss the implications for considering repeating and non-repeating FRBs for this fitted one-population model.

\subsection{FRB volumetric rate}


We find a local rate of bursts above $10^{39}$~erg and below a scattering of 10~ms at 600~MHz
of \volrate\ Gpc$^{-3}$ yr$^{-1}$.
This volumetric rate appears consistent with the rate presented by \citet{james+2021_sfr}, who found a rate of bursts above $10^{39}$~erg of $8.7^{+1.7}_{-3.9} \times 10^4$ bursts~Gpc$^{-3}$~yr$^{-1}$,
and is also broadly consistent with other studies as well \citep[e.g.,][]{lupiro2019, ravi2019repeatrates, luo2020lumin}.
On the surface, the consistent volumetric rates is interesting to note when considering the different frequency bands of the respective samples of FRBs.
However, great caution should be employed when directly comparing these values, as the rates presented by other studies do not employ a scattering cut.
As noted by \citet{catalog1}, there is evidence that there is a notable population of FRBs with a scattering timescale at 600~MHz larger than 10~ms that is missed by CHIME/FRB observations.
Although highly scattered bursts were excluded from this analysis due to the poorly constrained nature of their population,
including such highly scattered bursts may greatly increase our volumetric rate, possibly making the rate found in this analysis less comparable with other derived rates.

This volumetric rate of \volrate\ Gpc$^{-3}$ yr$^{-1}$ exceeds that of known populations of cataclysmic phenomena, especially when considering how the volumetric rate would increase when including highly scattered bursts.
As summarized by \citet{oguri2019}, Type Ia and core-collapse supernovae each have a local volumetric rate of $\mathcal{O}$(10$^4$)~Gpc$^{-3}$ yr$^{-1}$, with superluminous supernovae having a lower volumetric rate of $\mathcal{O}$(10$^2$)~Gpc$^{-3}$ yr$^{-1}$.
Both long and short gamma ray bursts have much lower local volumetric rates as well, with $\mathcal{O}$($\lesssim 10^2$)~Gpc$^{-3}$ yr$^{-1}$.
The inferred rate of binary compact object mergers (e.g., binary black hole, binary neutron star, and neutron star-black hole) with the cumulative Gravitational Wave Transient Catalog 3 (GWTC-3) are all $\mathcal{O}$($10^2$)~Gpc$^{-3}$ yr$^{-1}$ \citep{lvk2021_gwtc3}.
It has also been proposed that white dwarf mergers may occur in globular clusters in the local Universe at $\mathcal{O}$($10$)~Gpc$^{-3}$ yr$^{-1}$ \citep{kremer+2021_frbs_gcs}.



It is worth noting that when applying all the exclusion criteria in \ref{subsec:fiducial_distrs} \textit{except} for the cut on repeat bursts,
the Catalog~1 sample for populations analysis would have been 230 bursts instead of 225.
However, given the total net systemic uncertainty on the sky rate found by \citet{catalog1} was $^{+27\%}_{-25\%}$, this systemic uncertainty of treatment of repeaters is almost certainly a sub-dominant effect.
Thus, from a practical point of view, this volumetric rate can be interpreted as \textit{either} a rate of bursts \textit{or} of sources.
This reasoning is undeniably complicated by the fact that the observed rates are also strong functions of telescope sensitivity and the FRB repetition luminosity function.

Nonetheless, since it is possible that one-off bursts in our sample can actually be bursts from repeating sources, and the cut on repeaters is a sub-dominant effect in this analysis, if one looks purely at the obtained volumetric rate above 10$^{39}$~erg, it may be physically sensible to interpret it as a rate of bursts.
Extrapolating below the pivot energy introduces a large amount of uncertainty, and our model does not fit for a minimum energy, but it is reasonable to expect that the volumetric rate of FRBs is larger when including energies below $E_\mathrm{pivot}$.
If our observed volumetric rate is interpreted as a rate of bursts, 
as this rate exceeds that of known one-off event populations \citep[e.g.,][]{nicholl+2017}, 
such a scenario could suggest that FRBs are predominantly from repeaters,
most of which instruments such as CHIME/FRB may missed.
This in turn could imply that many apparent one-off FRBs are in fact repeat bursts from progenitors with long stretches of burst quiescence \citep{caleb+2021_review},
though the notion that all --- or even most --- FRBs are repeaters is one that has fallen out of relative favor \citep[e.g.,][]{ravi2019repeatrates, james+2020_which_frbs_repeat, ai+2021}.
There is also the possibility that repeaters and one-off FRBs are distinct populations with distinct populations parameter values; further discussion on this consideration can be found in Section~\ref{subsec:repeaters_v_nonrepeaters}.
Interpreting our observed volumetric rate as a rate of bursts can also suggest that FRBs predominantly come from a yet-unknown population of one-off events.
Nonetheless, it is important to emphasize that this discussion is driven purely by consideration of the numerical volumetric rate result;
our model $R(F,\dm|\bm \lambda)$ does not directly make any claims as to which progenitor origin scenario is preferred.
(The model does allow for source evolution, which could more directly argue for preferred progenitor scenarios based on evolution with star formation rate, but as discussed in Section~\ref{subsec:disc_sfr}, our results there are inconclusive.)

\subsection{FRB energy function}

\begin{figure}[htbp]
    \centering
    \includegraphics[width=\columnwidth]{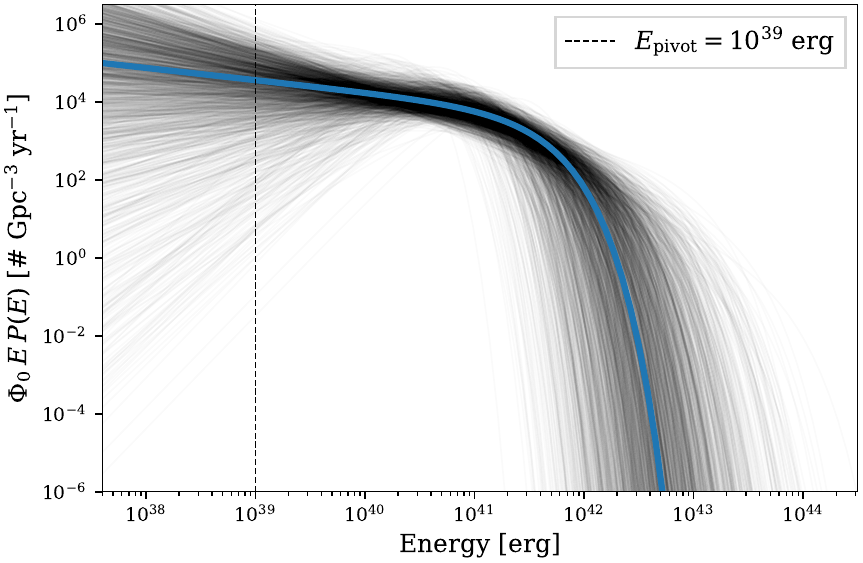}
    \caption{
        Plotted in gray are the results of the 
        allowable Schechter function values scaled by energy and the volumetric rate from
        the MCMC run, thinned by a factor of 500 for visualization purposes.
        The volumetric rate is quoted above the pivot energy, $E_\mathrm{pivot} = 10^{39}$~erg,
        which is overlaid with a dashed black line.
        The best-fit Schechter function from the MCMC run is overlaid in solid blue.
        Abundances at a given energy, above $E_\mathrm{pivot}$, can be directly read off the y-axis.
    }
    \label{fig:allowable_schechter}
\end{figure}

Constraining the energy function of FRBs is 
particularly illuminating for better understanding FRB origins.
Although we model the energy distribution as a Schechter function for all energies, we only quote a volumetric rate above a pivot energy $E_\mathrm{pivot} = 10^{39}$~erg.
We do not test for a minimum energy.
An allowable range of energy distributions is plotted in Figure~\ref{fig:allowable_schechter}.
We see that the shape of the function is quite well-constrained --- the characteristic energy after which the distribution exponentially falls off appears to visually consistently be on the order of $\approx$10$^{41}$~erg.
We find $\log_{10} E_\mathrm{char}$ (erg) to be $41.38^{+0.51}_{-0.50}$, in fairly good agreement with the \citet{james+2021_sfr} result of $\log_{10} E_\mathrm{max}$ (erg) of $41.70^{+0.53}_{-0.06}$.
(Note \citet{james+2021_sfr} preferred to model the energy distribution with a power law with a maximum energy cutoff rather than as a Schechter function, although both models were tested with comparable results.)

We also find a differential power-law energy distribution index of $\gamma =$~\gammaval.
This result is in broad agreement with the results of \citet{hashimoto2022}, who also used Catalog~1 data to explore the energy function of FRBs,
albeit employing a different methodology (e.g., they do not fit for a distance distribution simultaneously) and using a different subset of bursts released by \citet{catalog1}.
Within error bars, this power-law index is also in agreement with what \citet{luo2020lumin} found ($-1.79^{+0.31}_{-0.35}$) using a heterogenous sample of FRBs, including bursts from UTMOST-SS \citep{caleb+2017} and from CRAFT \citep{shannon2018}.
Interesting to note is that \citet{james+2021_sfr}, who developed the model $R(F,\dm | \bm \lambda)$ we use in this analysis,
find a steeper power-law index value ($-2.09^{+0.14}_{-0.10}$) using a sample of ASKAP and Parkes FRBs,
though it is still consistent with our results within 2$\sigma$ statistical uncertainties.
Further data and larger numbers of FRB observations are needed to explore the consistency of FRB energy distributions across different frequency bands.

\subsection{FRB spectral dependence with $\alpha$}

When interpreting $\alpha$ as a true spectral index, every FRB is observed as a broad-band burst with the same spectral index.
Using a sample of 23 FRBs detected with ASKAP, \citet{macquart+2019_spectralprops} find $\alpha = -1.5$, a default value adopted in other studies \citep[e.g.,][]{lupiro2019}.
Using a Gaussian prior on $\alpha = -1.5 \pm 0.3$, \citet{james+2021_zDM} find a consistent $\alpha = -1.55^{+0.18}_{-0.18}$.
Our result of $\alpha=$~\alphaval\ is consistent with
$\alpha \approx -1.5$, although our statistical uncertainties are larger than those of the aforementioned quoted values.

The treatment of $\alpha$ is far from standardized in the literature.
Multiple studies fix $\alpha = 0$ \citep[e.g.,][]{caleb+2016, luo2020lumin, chawla2022} when studying their sample.
In particular, \citet{chawla2022} set $\alpha = 0$ for the Catalog~1 sample
in order to be consistent with the implicit assumptions made when measuring fluences for Catalog~1 bursts \citep{catalog1}.
Also, as has been detailed by \citet{james+2021_zDM} and mentioned by \citet{law+2017_frb121102},
there is little meaning of a ``spectral index'' for an individual FRB with a limited band occupancy.
Instead, when interpreting $\alpha$ as a frequency-dependent rate, every FRB is observed as a narrow-band burst, depending on the frequency-dependent survey sensitivity.
Our rate interpretation result is $\alpha' = $~\alphavalrateinterp, which is slightly flatter than our ``spectral index'' interpretation of $\alpha =$~\alphaval\ but again, not significantly so --- especially when considering the uncertainties.
Certainly more work remains to better constrain the spectral index of FRBs, and how best to treat the systemic uncertainty of how to interpret $\alpha$.

\subsection{FRB evolution with star formation history}
\label{subsec:disc_sfr}


The progenitors of FRBs are a widely debated topic, and one way to probe their origins as a population is to track their redshift distribution with respect to star formation history.
If FRBs arise predominantly from magnetars, as the association of the FRB-like burst with the Galactic magnetar SGR~1935+2154 might suggest \citep{chimefrb_sgr1935, bochenek+2020_stare2_sgr1935}, then they would closely track the star formation history (SFH) of the universe.
On the other hand, if FRBs arise predominantly from older stellar populations, as the association of FRB~20200120E with a globular cluster in M81 might suggest \citep{bhardwaj+2021_m81, kirsten+2022_m81}, then they would not appear to trace the SFH.
There is also the possibility that FRBs arise from younger and older stellar populations in comparable proportions \citep[e.g.,][]{guo_wei_2022_frbsubclasses}, but given the limited information regarding FRB host environments and the risk of over-interpreting available data, we do not consider this scenario in this work.

\begin{figure}[htbp]
    \centering
    \includegraphics[width=\columnwidth]{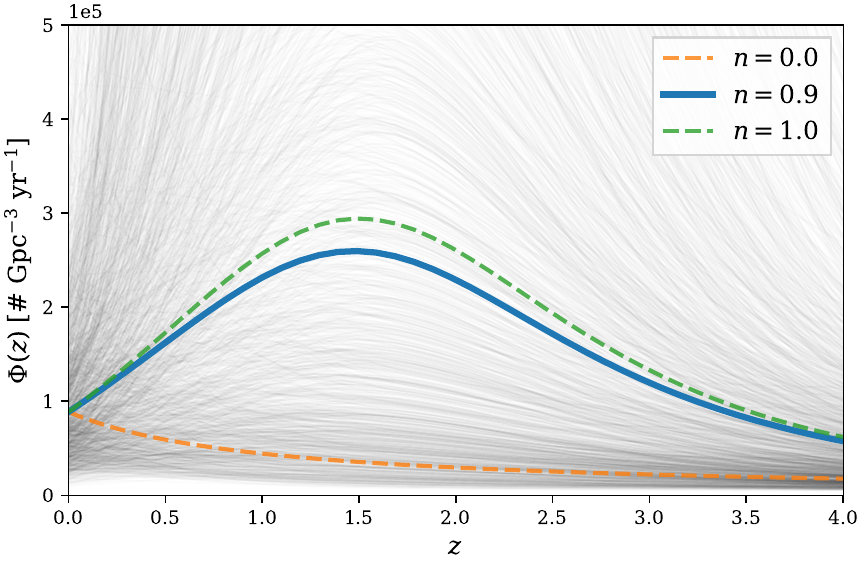}
    \caption{
        Results of how the Catalog~1 FRBs evolve with SFR, $\Phi(z)$.
        Gray lines denote allowable behavior of $\Phi(z)$ using parameters from the MCMC run, thinned by a factor of 500 for visualization purposes.
        The best-fit $\Phi(z)$ from the MCMC is overlaid in solid blue.
        The lower dashed orange curve represents no evolution with SFR, and the upper dashed yellow curve represents perfect evolution with SFR.
    }
    \label{fig:allowable_sfr}
\end{figure}

Instead, the formalism of this population model simply aims to quantify how closely the FRB population evolves with SFR.
\citet{james+2021_sfr} claim that the FRB population does evolve with SFR, a result that is not incompatible with our results of $n =$~\n\ when considering the different samples of FRBs used.
The frequency coverage differs between Parkes (1157 -- 1546~MHz) and ASKAP (1129 -- 1465~MHz) compared to CHIME (400 -- 800~MHz), raising the possibility that there may be frequency-dependent evolution behavior.
However, there is no concrete evidence in our analysis to support this claim, and again, any related investigation is outside the scope of this work.
Other studies have investigated the cosmic evolution of FRBs \citep[e.g.,][]{hashimoto2022, zhangzhang22, qiang2022} using Catalog~1 data, and have concluded that FRBs from the CHIME/FRB Catalog~1 sample do not evolve with SFR.
Our result of $n =$~\n\ is much less conclusive; the large uncertainties on our results are also visualized in Figure~\ref{fig:allowable_sfr}.
Of our $10^6$ chains, $\approx$7.2\% give $n \leq 0$ (the rest of the parameter results vary across chains; there is no real physical interpretation for $n<0$, but we allowed the range for completeness).
Our result under the ``rate interpretation'' of $\alpha$ also gives a poorly constrained $n'=$~\nrateinterp;
of those chains, $\approx$4.5\% give $n \leq 0$.
We note that under the ``rate interpretation'' of $\alpha$, our results tend towards stronger evolution with SFR --- consistent with what \citet{james+2021_zDM} found.
However, the statistical uncertainties on both parameters, in both interpretations, are too large to draw any significant conclusions about how the interpretation of $\alpha$ may affect the modeling of FRB evolution with SFR.

The difference between the results of this study and of previous studies using Catalog~1 data can be explained in a few ways.
Using the full injections sample, this analysis fits a population model which simultaneously models both the energy distributions and the redshift distributions of FRBs.
In the work presented by \citet{hashimoto2022}, redshifts are directly inferred from $\dm$ measurements rather than treating them as separate parameters, and they thus explored the redshift evolution of the derived FRB energy functions.
Both \citet{qiang2022} and \citet{zhangzhang22} generate Monte Carlo distributions from varied redshift and energy distributions, and then test them against the publicly available CHIME/FRB Catalog~1 data.
While this approach allows for explicit testing of different evolutionary models (i.e., including time delays), it does not fit specific parameters to the data --- it can only test how consistent a dataset may be with a tested distribution.
Additionally, these previous studies did not have access to the full injections sample to calibrate out observational selection effects with their methodology.
We do test one more toy model, elaborated on in Appendix~\ref{sec:time_delay_sfrmodel}, to explore the potential constraining power of our data for FRBs from older stellar populations.
Ultimately, though, because our result of $n =$~\n\ has such large uncertainties, we conclude that our data cannot meaningfully constrain how the FRB population evolves with the star formation history of the universe.
However, localizations to host galaxies and direct redshift measurements for the FRBs may enable better constraints in the future.
They would enable determination of information such as stellar masses,
which one can then use to also test whether FRB host galaxies track the stellar mass function (e.g., as \citealt{bhandari2022} did with ASKAP FRBs).

\subsection{Host galaxy DM contribution}

\begin{figure}[htbp]
    \centering
    \includegraphics[width=\columnwidth]{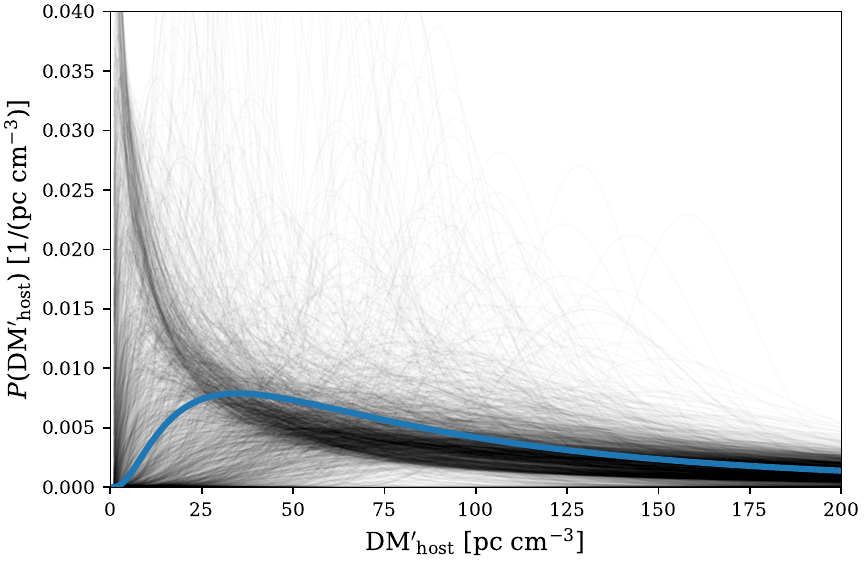}
    \caption{
        Plotted in gray are the results of the allowable distributions of the rest-frame contribution to the $\dm$ from the host galaxy of the FRB, $P(\dm_\mathrm{host}\sp{\prime})$, from the MCMC run, thinned by a factor of 500 for visualization purposes.
        The best-fit distribution from the MCMC run is overlaid in solid blue.
        Due to the extended tail nature of this distribution, only values up to $\dm_\mathrm{host}\sp{\prime}=200$~pc~cm$^{-3}$ are plotted.
        However, higher $\dm_\mathrm{host}\sp{\prime}$ values are certainly allowed.
    }
    \label{fig:allowable_dmhost}
\end{figure}

The rest-frame host galaxy contribution to the $\dm$ of an FRB was modeled with a log-normal distribution,
parameterized by $\mu_\mathrm{host}$ and $\sigma_\mathrm{host}$.
However, the values of $\mu_\mathrm{host}$ and $\sigma_\mathrm{host}$ do not directly
correspond to physical values.
Instead, the median and standard deviation of the log-normal rest-frame $\dm_\mathrm{host}$ distribution described by those parameters are \medhost~pc~cm$^{-3}$ and \stddevhost~pc~cm$^{-3}$, respectively.
As illustrated by the statistical uncertainties, these parameters are not well-constrained;
how poorly constrained these parameters are can also be visualized with the allowable
$P(\dm_\mathrm{host}\sp{\prime})$ distributions plotted in Figure~\ref{fig:allowable_dmhost}.
As such, caution must be employed when comparing these values against those found in the literature.
Nonetheless, we highlight a few observations.

\citet{macquart+2020_dmz} and \citet{james+2021_sfr, james+2021_zDM} also generically modeled the intrinsic host galaxy contribution to the $\dm$ with a log-normal distribution,
with only the latter study explicitly quoting derived values.
In the work presented by \citet{james+2021_sfr}, they quote a log-mean host $\dm$ contribution of $129^{+66}_{-48}$~pc~cm$^{-3}$ on top of a typical Galactic contribution of $\dm_\mathrm{disk} + \dm_\mathrm{halo} = 50 + 35 =$ 85~pc~cm$^{-3}$.
It is important to note that \citet{james+2021_sfr} used a different sample of FRBs from different surveys (at different observing frequencies), did not apply a scattering cut for $\tau > 10$ms to their sample as we did, and
also quote a typical Galactic contribution of 85~pc~cm$^{-3}$ instead of $=$80~pc~cm$^{-3}$, although this latter difference is likely to be minimal (Section~\ref{sec:varyDML}).
However, in Figure~\ref{fig:mcmc_corner}, we can see that our MCMC chains also strongly disfavor low $\sigma_\mathrm{host}$ and high $\mu_\mathrm{host}$ values, a finding consistent with \citet{james+2021_sfr}.

Both \citet{james+2021_sfr} and our analysis find ``central'' values of $\dm_\mathrm{host}'$ well within statistical errors of each other, as well as within
``typical'' host galaxy contributions which are often quoted as $\sim$50~pc~cm$^{-3}$ \citep[e.g.,][]{shannon2018} or $\sim$100~pc~cm$^{-3}$ \citep[e.g.,][]{cordes_chatterjee2019_ARAA}.
One should also note that the log-normal nature of the distribution allows for the occasional extraordinarily high host galaxy $\dm$ such as the estimated $\dm_\mathrm{host} \approx 903^{+72}_{-111}$~pc~cm$^{-3}$ found for FRB 20190520B by \citet{niu+2021_121102twin}.
This is also in agreement with \citet{rafieiravandi2021} finding statistical evidence for some FRBs having a large host $\dm$ (defined therein as $\sim$400~pc~cm$^{-3}$).
Overall, we note that the poor constraints we obtain with the Catalog~1 data are perhaps unsurprising when one considers that no redshift information was used for any FRBs in this study.


\subsection{Considering repeaters vs non-repeaters}
\label{subsec:repeaters_v_nonrepeaters}

A growing number of FRBs has been observed to repeat \citep[e.g.,][]{spitler+2016_frb121102, scholz+2016_frb121102, RN1, R3, RN2, bhardwaj+2021_m81, lanman+2021_R67}.
There is currently no reliable method to identify
which FRBs are from repeaters,
but the growing sample of observed repeat bursts has led to some promising methods.
\citet{pleunis+2021_morphology} found that repeaters could be a distinct population from apparent one-offs based on the temporal and spectral studies of the Catalog~1 sample
Also using the Catalog~1 sample, \citet{chen+2022_repeatingfrbML} have recently used unsupervised machine learning
to create a catalog of repeater source candidates.
Of course, all this is assuming that there are two distinct populations of FRB sources: those that repeat, and those that do not.
Although observationally such a classification scheme is intuitive,
it is also possible that all FRBs repeat, or all FRBs
represent distinct phases of the same class of progenitors
\citep[e.g.,][]{caleb+2019_doallfrbsrepeat, pleunis+2021_morphology, cui+2021}.

If most FRBs are from repeaters, the results presented here are likely to be affected to a stronger degree than if a lower fraction of FRBs is from repeaters.
As mentioned in Section~\ref{subsec:fiducial_distrs}, to minimize unquantifiable bias, we include only the first-detected burst from any observed FRB source.
However, one should expect the repetition rate of FRBs to affect distributions of observed parameters.
Naively, the presence of increasing numbers of repeating FRBs
would cause the burst rate within an observed volume of the universe
to be increasingly greater than expected.
Likewise, \citet{gardenier+2021_repeat} noted that
repeating FRBs should have a lower mean $\dm$ than apparent one-off FRBs,
although \citet{catalog1} found that the distributions of repeating FRBs and apparent one-off FRBs are consistent with being drawn from the same underlying $\dm$ distribution.
\citet{pleunis+2021_morphology} show that there are observable differences in the spectral parameters (e.g., $\alpha$) of repeaters and non-repeaters,
and of course the differences could extend to physically different values of
$\dm_\mathrm{host}$ contributions (e.g., $\mu_\mathrm{host}$ and $\sigma_\mathrm{host}$), evolution with SFH (e.g., $n$), and more.

Although it is tempting to conduct a populations study
considering only apparent non-repeaters from the Catalog~1 sample,
it is likely that some apparent one-off bursts are actually from repeater sources, with both \citet{pleunis+2021_morphology} and \citet{chen+2022_repeatingfrbML}
proposing candidate apparent one-off bursts that may actually be from repeater sources.
In fact, FRB20190304A was identified as an apparent non-repeater source in Catalog~1, but since then another burst has been observed from this source (CHIME/FRB~Collaboration et al., private communication), establishing its nature as a repeater instead.
There is also the fact that CHIME/FRB has only observed one burst from the first repeating source FRB~121102 \citep{josephy+2019_R1}, despite there being many more additional observed bursts present in the literature
\citep[e.g.,][]{rajwade+2020_frb121102_periodicity, li+2021_FASTfrb121102}.
Furthermore, \citet{good+2022_chimefrbarecibo} demonstrate that the possibility of a low-repetition sub-population may further complicate the observational identification of repeater FRBs.
The presence of such as-yet-unidentified repeat bursts could represent a not insignificant ``contamination'' of the nominal non-repeating sample of FRBs.
\citet{chen+2022_repeatingfrbML} suggest a large potential ``contamination'' of repeaters,
but \citet{pleunis+2021_morphology} warn that instrumental effects, such as chromatic sensitivities away from the detection beam centers,
make it difficult to isolate the true repeaters.
This is because such instrumental effects would result in observed bursts with biased low bandwidths that also cause mixing of bursts with spectra that are intrinsically broad and narrow.
Further complicating this unquantifiable effect of as-yet unidentified repeaters in a sample of nominally one-off bursts is
that the injections framework of CHIME/FRB currently only injects bursts from a single population distribution of FRB parameters.
Nonetheless, as larger samples of well-defined repeaters are identified, we expect our ability to
do comparative populations analyses between observed
repeat bursts and one-off bursts will improve.

\section{CHIME/FRB Outrigger Predictions}
\label{sec:outriggers}

For the majority of FRBs detected by CHIME/FRB, the localization precision is insufficient for identifying host galaxies for these bursts.
Host galaxies, and thus redshifts, for FRBs are highly desirable for both understanding the nature of FRB origins \citep{bhandari2022} and using them as cosmological probes \citep{masuisigurdson2015}.
Enabling larger-scale localization abilities is a high-priority science goal for FRBs, and one that will be addressed with the CHIME/FRB~Outriggers project \citep[e.g.,][]{leung+2021_vlbi, cary+2021_rnaas, menaparra+2022_clockoutriggers, cassanelli+2022_aro}.
Through CHIME/FRB~Outriggers, CHIME-like outrigger telescopes will be deployed at continental baselines to form a very-long-baseline interferometry (VLBI) network.
While CHIME/FRB~Outriggers will trigger on a subset of sources observed by CHIME/FRB (e.g., not on low--$\snr$ bursts),
the VLBI network will still obtain near real-time, sub-arcsecond localization capability for hundreds of FRBs a year \citep{menaparra+2022_clockoutriggers}.
Although it is unrealistic to expect that every FRB will have its VLBI position followed-up with optical observations to obtain spectroscopic redshifts,
it would certainly be possible for spectroscopic redshifts to be obtained for a much more significant fraction of FRBs than is currently possible.
Additionally, a larger fraction of FRBs may have reasonably reliable photometric redshifts obtained by cross-matching the VLBI localizations with photometric galaxy surveys such as the Dark Energy Spectroscopic Instrument (DESI) Legacy Imaging Surveys \citep{dey+2019}.
While the photometric redshifts would not be as reliable as the spectroscopic redshifts,
they would be more informative than redshifts inferred from $\dm$ values.
In any event, further additional redshift information would in turn allow for much more sensitive constraints on the intrinsic $\dm_\mathrm{host}$ contribution.

With the methodology used in this paper, we can obtain a prediction for the observed FRB redshift distribution that CHIME/FRB~Outriggers may see.
Recall Equation~\ref{eqn:rate_fdm_distr}:
\begin{equation*}
	 R(F, \dm | \bm{\lambda}) = \int dz R(F, z | \bm{\lambda_1}) P(\dm | z, \bm{\lambda_2}).
\end{equation*}
It is here that the model $R(F, \dm | \bm \lambda)$ is marginalized over $z$.
As described in Section~\ref{sec:methodology}, $R(F, \dm | \bm \lambda)$ is used to construct parameter-dependent weights, which are then used to create a model prediction $\zeta_{ij}$, a weighted histogram of the synthetic data in $\snr$ and $\dm$ (Equation~\ref{eqn:mu_ij}).
Properly normalized, this earlier obtained $\zeta_{ij}$ can be translated to an observed rate distribution of $\snr$ and $\dm$, given parameters $\bm \lambda$, over the redshift range from $1 \times 10^{-3}$ to $4$.

Instead of integrating over $z$ and ending up with a 2D model prediction (histogram) in $\snr$ and $\dm$,
we can construct a 3D model prediction that depends on $\snr$, $\dm$, and $z$:
\begin{eqnarray}
    \label{eqn:mu_ij_def}
    \zeta_{ij} {\Bigr |}_{z_\mathrm{val}}  
    &=& \sum_{m=1}^{n_{ij}} W_m(F, \dm, z_\mathrm{val}, \mathbf{\Theta}, \bm \lambda)\\
    &=& \sum_{m=1}^{n_{ij}} \frac{R(F, \dm, z_\mathrm{val} | \bm \lambda) \Delta t }{R_\mathrm{init}(F, \dm) }
\end{eqnarray}
where
\begin{equation}
    \label{eqn:R_F_DM_z_def}
    R(F, \dm, z_\mathrm{val} | \bm{\lambda}) = \Delta z \; R(F, z_\mathrm{val} | \bm{\lambda_1}) P(\dm | z_\mathrm{val}, \bm{\lambda_2}),
\end{equation}
and $\Delta z$ is estimated by the midpoint spacing between the evaluated $z_\mathrm{val}$ points across the full redshift range range from $1 \times 10^{-3}$ to $4$.
(Note that for simplicity of notation, $\mathbf{\Theta}$ was dropped as their distributions were held to the fiducial models found in Catalog~1 for all evaluated models, for each burst.)

\begin{figure}[htbp]
    \centering
    \includegraphics[width=\columnwidth]{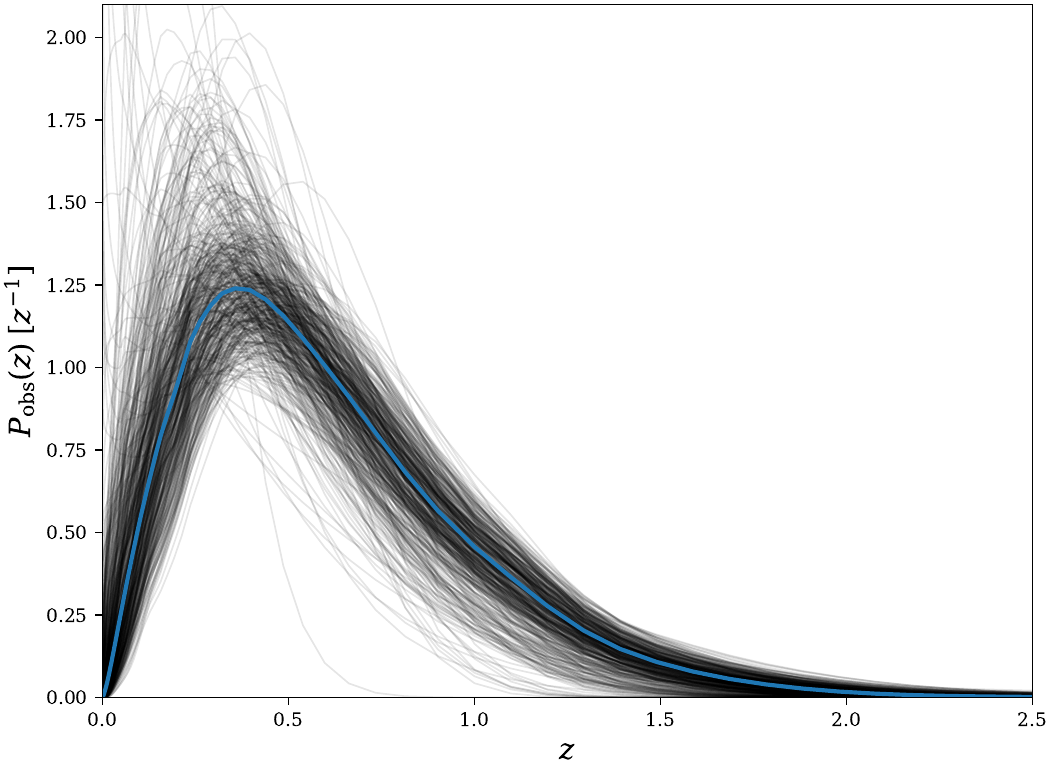}
    \caption{
        Plotted in gray are the results of the allowable observed FRB redshift distributions $R_\mathrm{obs}(z)$ from the MCMC run, thinned by a factor of 2000 for visualization purposes.
        The $R_\mathrm{obs}(z)$ prediction using the best-fit distribution from the MCMC run is overlaid in solid blue.
        Although the model was fit up to $z=4$, only redshifts up to $z=2.5$ are plotted.
    }
    \label{fig:allowable_Pobsz}
\end{figure}

This 3D histogram, after dividing by the bin widths in $z$, gives the prediction for the observed CHIME/FRB Catalog~1 $\dm$--$\snr$--$z$ distribution per $z$ for a given model $R(F, \dm | \bm \lambda)$.
We can thus sum over the $\snr$ bins (which span the values 12--200) 
and the $\dm$ bins (which span the values 100--3500~pc~cm$^{-3}$) to obtain the prediction for the observed redshift distribution, $P_\mathrm{obs}(z)$.
The range of predicted redshift distributions our MCMC run allows appears reasonably well-constrained, and is shown in Figure~\ref{fig:allowable_Pobsz}.

The $P_\mathrm{obs}(z)$ distribution predicted by the best-fit model peaks at around a redshift of $z \approx 0.36$.
This result is consistent with the $\dm$ distribution observed by CHIME/FRB shown in \citet{catalog1}, where the peak of that distribution is around $\dm \approx 500$~pc~cm$^{-3}$.
That the redshift peak is below $z \sim 0.5$ suggests that there is a noticeable fraction of low to intermediate-$z$ FRBs that have a sizable $\dm$ contribution from astrophysical environments other than the cosmological intergalactic medium (e.g., local environments, host galaxies, or intervening galaxies).
This $P_\mathrm{obs}(z)$ prediction also contains $\approx$99.4\% of its area within $z\leq2$,
suggesting that CHIME/FRB~Outriggers will see a negligible number FRBs with $z>2$.
Such a result implies that the sample of FRBs observed by CHIME/FRB~Outriggers will be limited in its ability to be used as high-$z$ cosmological probes \citep[e.g.,][]{linder2020_frb_He_reionization}.
It is worth noting this prediction is based on the sample that survived the statistical cuts detailed in Section~\ref{subsec:fiducial_distrs};
therefore, it is likely that the true $R_\mathrm{obs}(z)$ distribution will differ when including observed bursts that do not pass these cuts (such as bursts with $\dm < 1.5$ max($\dm_\mathrm{NE2001}, \dm_\mathrm{YMW16}$) and $\snr < 12$).
Nevertheless, this prediction supports the notion that low-$z$ cosmological studies such as lensing from foreground structures will likely be a promising scientific application of FRBs in the near future \citep[e.g.,][]{kader+2022_lens, leung+2022_lens}.

\begin{figure*}[htbp]
    \centering
    \includegraphics[width=\textwidth]{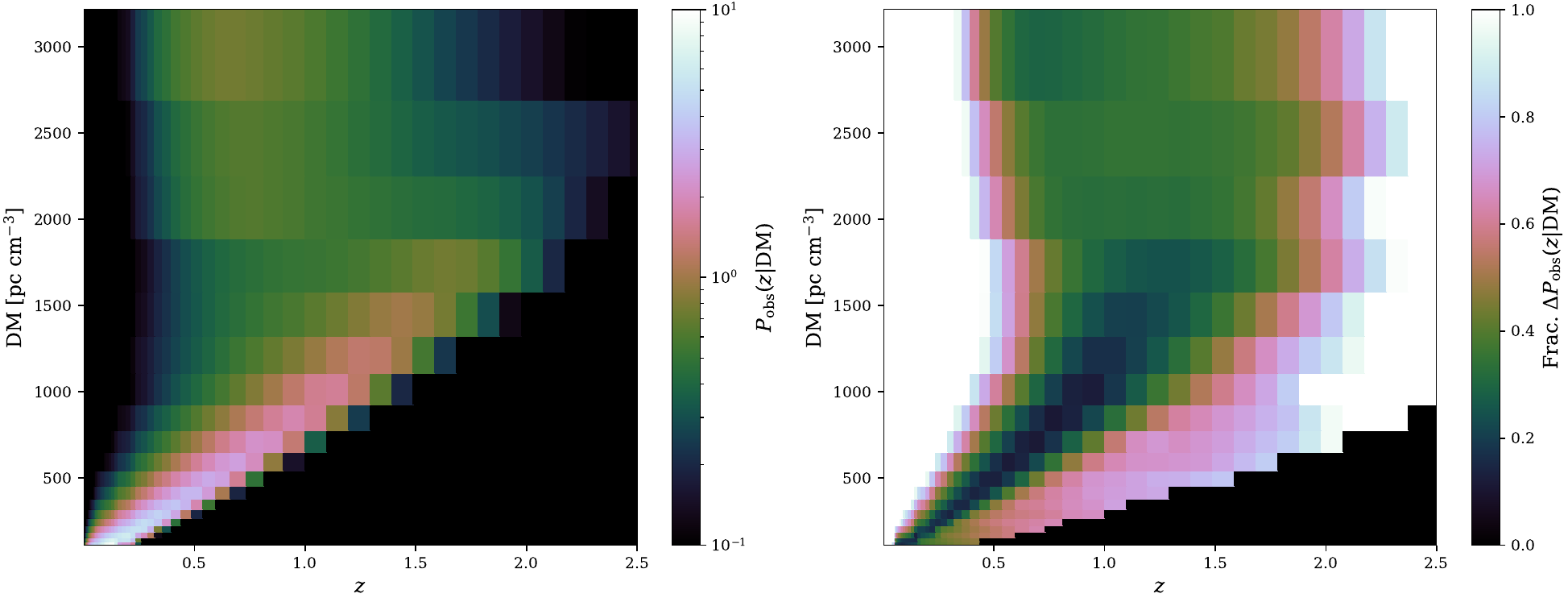}
    \caption{
        \textbf{Left panel:} The median $P_\mathrm{obs}(z|\dm)$ distribution for the results of the MCMC run, thinned by a factor of 2000 for computational speed purposes.
        Evidence for a ``turnover'' of the DM--$z$ relation can be seen at higher DMs and lower redshifts,
        although most of the FRBs appear to occur at $\dm < 2000$~pc~cm$^{-3}$ and $z<2$.
        \textbf{Right panel:} The standard deviation of the of the $P_\mathrm{obs}(\dm, z)$ distributions at each point divided by the median $P_\mathrm{obs}(\dm, z)$ (shown on the left) to visualize the fractional uncertainty of $P_\mathrm{obs}(z|\dm)$ values across MCMC realizations.
    }
    \label{fig:pred_Pobsz_givenDM}
\end{figure*}

We can also take the 3D histogram that gives the prediction for the observed CHIME/FRB Catalog~1 $\dm$-$\snr$-$z$ distribution, given by Equations~\ref{eqn:mu_ij_def} and \ref{eqn:R_F_DM_z_def}, and divide by the bin widths in $z$ and $\dm$.
Then, we can sum over the $\snr$ bins and normalize the distribution such that we obtain a prediction for the observed joint redshift and $\dm$ distribution, per $z$ and per $\dm$: $P_\mathrm{obs}(z|\dm)$.
We show a visualization of what $P_\mathrm{obs}(z|\dm)$ predictions are allowed by the MCMC run in Figure~\ref{fig:pred_Pobsz_givenDM}.

In the left-hand panel of Figure~\ref{fig:pred_Pobsz_givenDM},
the $P_\mathrm{obs}(z|\dm)$ result shows that, although the relation between the redshift and $\dm$ of FRBs is a reasonably tight, single-modal distribution \citep[e.g.,][]{shannon2018, macquart+2020_dmz}, for a survey sample as large as Catalog~1, there is a ``turnover'' at high $\dm$ where the true redshift of the FRB is much lower than would be expected by applying a naive $\sim$linear DM--$z$ relation \citep[first shown by][]{james+2021_zDM}.
This can be seen the apparent excess of probability in the overlapping region of $0.5 \lesssim z \lesssim 1.0$ and $\dm \gtrsim 2000$~pc~cm$^{-3}$.
There, the plot is saying that, for an FRB observed by CHIME/FRB~Outriggers with a measured $\dm \gtrsim 2000$~pc~cm$^{-3}$, it is \textit{more} likely for the source to have a redshift $z<1$ with a high $\dm_\mathrm{host}$ contribution than a redshift $z>2$ with a large $\dm_\mathrm{IGM}$ contribution.
This finding is consistent with \citet{rafieiravandi2021}, who find statistical evidence for a nonzero correlation between high-$\dm$ FRBs ($\geq 785$~pc~cm$^{-3}$) and galaxies at $z \sim 0.4$, with a $\dm_\mathrm{host}$ contribution $\sim$greater than the $\dm_\mathrm{IGM}$ contribution.
This result once again emphasizes the importance of actually obtaining redshifts for FRBs rather than relying on inferred redshift values when doing cosmological studies,
especially for high-DM FRBs.
Of course, one must keep in mind that these predictions are for an overall population, and that an individual detected FRB can certainly be a cosmologically distant source with high intrinsic luminosity.
In fact, \citet{ryder+2022_frbz1} recently reported the discovery of FRB 20220610A, the first FRB observed at $z \gtrsim 1$.

The right-hand panel of this figure shows fractional uncertainties associated with our prediction for $P_\mathrm{obs}(z|\dm)$ across the chains in our MCMC run.
This plot illustrates that, along the relation between $z$ and $\dm$, the predicted $P_\mathrm{obs}(z|\dm)$ is quite consistent across realizations.
The further the deviation from that relation, the more ``scatter'' there is between predictions in the run.
It is also worth noting that ``fractional uncertainty'' can also mean ``scatter surrounding how close this value is to zero across runs,''
as appears to be the case in the region with $\dm < 500$~pc~cm$^{-3}$ and $0.5 < z < 1.0$ --- a region that appears to have negligible probability in the left-hand panel of this figure, but can have scatter due to variations in large scale structure noise.

\section{Conclusion}
\label{sec:conclusion}

Using the CHIME/FRB Catalog~1 data, we improve upon the initial Catalog~1 populations analysis by fitting a more physical model to the fluence and DMs of the FRBs,
a model initially developed in previous works \citep[e.g.,][]{luo2020lumin, james+2021_zDM}.
We use the weighting formalism developed in \citet{catalog1}, which makes use of the injected synthetic pulses to calibrate out selection function effects when fitting the model to the observed FRB data.
The model fits for 7 parameters: a volumetric rate $\Phi_0$, a spectral index $\alpha$, a differential power-law energy index $\gamma$, a characteristic exponential cutoff energy $E_\mathrm{char}$, a power-law index $n$ for smooth-scaling evolution with SFR, and parameters $\mu_\mathrm{host}$ and $\sigma_\mathrm{host}$ that parameterize the log-normal distribution of the host $\dm$ contribution.
We also explore the ``rate interpretation'' of $\alpha$.

We find a volumetric rate of FRBs above $10^{39}$~erg and below a scattering of 10~ms at 600~MHz of $\Phi_0 =$~[\volratenum(stat.)$^{+2.0}_{-1.8}$(sys.)]$\times 10^4$~Gpc$^{-3}$~yr$^{-1}$.
We also find tight constraints on the energy distribution observed by CHIME/FRB, modeled as a Schechter function, with $E_\mathrm{char} =$~\Echar~erg and $\gamma =$~\gammaval.
However, the constraints on the remainder of the parameters are much weaker.
In particular, the constraints on $n$, $\mu_\mathrm{host}$, and $\sigma_\mathrm{host}$ may significantly improve with the use of redshift information for many FRBs.
Such a sample of redshift information would be enabled by the upcoming CHIME/FRB~Outriggers project, for which we predict the observed FRB redshift distribution.
The distribution shows that the application of CHIME-detected FRBs to cosmological studies may be limited to the $z<2$ regime, which seems to limit the possibility of using these FRBs to probe events such as
Helium reionization, but does allow for FRB lensing studies.

As the amount of data seems to be the limiting factor for this FRB populations study, it is promising that CHIME/FRB has continued to detect a large number of FRBs since Catalog~1 (which only included bursts up to 2019 July 1).
These observed FRBs, combined with knowledge about survey selection effects provided by the injections system, and possibly redshift information with the commissioning of CHIME/FRB~Outriggers, suggest much more powerful populations-level constraints in the near future.
With more data comes more responsibility to handle subtle effects --- namely, how best to treat repeating sources versus apparent one-off bursts when modeling the FRB distribution.
Such a question would be of particular importance when exploring population behavior in a parameter space where repeater bursts would dominate, such as events with large pulse widths.
In addition, proper modeling of repeater and non-repeater sources --- whether apparently or intrinsically non-repeating --- would enable future large-sample population studies to investigate to what extent all FRBs may repeat.
Further in the future, care must be taken to properly incorporate redshift information into populations studies while accounting for the selection biases (e.g., repeater or high-$\snr$) that lead to some, but not all, FRBs having associated redshifts.

\section*{acknowledgments}
We acknowledge that CHIME is located on the traditional, ancestral, and unceded territory of the Syilx/Okanagan people. We are grateful to the staff of the Dominion Radio Astrophysical Observatory, which is operated by the National Research Council of Canada.  CHIME is funded by a grant from the Canada Foundation for Innovation (CFI) 2012 Leading Edge Fund (Project 31170) and by contributions from the provinces of British Columbia, Qu\'{e}bec and Ontario. The CHIME/FRB Project is funded by a grant from the CFI 2015 Innovation Fund (Project 33213) and by contributions from the provinces of British Columbia and Qu\'{e}bec, and by the Dunlap Institute for Astronomy and Astrophysics at the University of Toronto. Additional support was provided by the Canadian Institute for Advanced Research (CIFAR), McGill University and the McGill Space Institute thanks to the Trottier Family Foundation, and the University of British Columbia.
A.B.P. is a McGill Space Institute (MSI) Fellow and a Fonds de Recherche du Quebec - Nature et Technologies (FRQNT) postdoctoral fellow.
\allacks

\appendix
\restartappendixnumbering

\section{Details of the Fluence--DM Distribution Model}
\label{sec:modeldeets}

\subsection{Joint fluence and redshift distribution}
\label{sec:R(F,z)}
In this section, we focus on how to obtain $R(F, z | \bm{\lambda_1})$.
The parameter vector $\bm{\lambda_1}$ contains information related to FRB energetics.

First, we define the energy function $ P(E) $.
We model $ P(E) $ with a Schechter function with a characteristic exponential cutoff energy $E_\mathrm{char}$ and power-law index $\gamma$, e.g.,
\begin{equation}\label{eqn:schechter_appendix}
	P(E) dE \propto \frac{1}{E_\mathrm{char}} \left(\frac{E}{E_\mathrm{char}}\right)^\gamma \exp{\left[-\frac{E}{E_\mathrm{char}}\right]} dE.
\end{equation}

In order for $ \int  E \: P(E) \: dE $ in Equation~\ref{eqn:schechter_appendix} to not diverge, we must have $ \gamma < -2 $.
However, this function is not truly normalizable, i.e., we cannot guarantee $ \int_0^\infty  P(E) \: dE = 1 $ for all redshifts.
Thus, we shall choose to normalize the function such that $ \int_{E_\mathrm{pivot}}^\infty  P(E) \: dE = 1 $
while assuming the energy itself extrapolates to follow the Schechter function.
This allows us to assume this energy distribution is valid for all FRB energies down to a minimum (but unknown) energy $E_\mathrm{min}$, while quoting a rate of FRBs above $ E_\mathrm{pivot} $.
For a reasonable choice of $ E_\mathrm{pivot} $,  $\Phi_0$ should be weakly correlated with $ \gamma $.
We chose a pivot energy of $10^{39}$~erg as it is above the threshold of ruled out minimum FRB energies \citep[e.g.,][]{james+2021_zDM};
in our results in Figure~\ref{fig:mcmc_corner}, we can see that $\Phi_0$ and $ \gamma $ are clearly correlated, but because their correlation isn't strong, we did not fine tune $ E_\mathrm{pivot} $ further so as to have $\Phi_0$ and $ \gamma $ be more weakly correlated.

To normalize this energy distribution, recall that the integral of a Schechter function is an upper incomplete gamma function, defined as
\begin{equation}
    \Gamma \left(\gamma+1, \frac{E_\mathrm{pivot}}{E_\mathrm{char}} \right)
    =
    \int_{E_\mathrm{pivot}}^\infty \frac{1}{E_\mathrm{char}} \left(\frac{E}{E_\mathrm{char}}\right)^\gamma \exp{\left[-\frac{E}{E_\mathrm{char}}\right]} dE.
\end{equation}

Thus, the ``normalized'' Schechter function above a pivot energy $E_\mathrm{pivot}$ is
\begin{equation}
    P(E) dE = \frac{1}{\Gamma(\gamma+1, E_\mathrm{pivot} / E_\mathrm{char})}
    \frac{1}{E_\mathrm{char}} \left(\frac{E}{E_\mathrm{char}}\right)^\gamma \exp{\left[-\frac{E}{E_\mathrm{char}}\right]} dE.
\end{equation}
With this normalization, all rates should be interpreted as a rate above the pivot energy $E_\mathrm{pivot}$.

We can convert between the FRB energy and fluence with
\begin{equation}\label{eqn:E(F)}
	E = \frac{4 \pi D^2_L(z)}{(1+z)^{2+\alpha}} \Delta \nu F,
\end{equation}
where $E$ is the FRB energy, $D_L(z)$ is the luminosity distance, $\alpha$ is the spectral index ($F \propto \nu^\alpha$), and $\Delta \nu$ is the frequency bandwidth (taken as 1 GHz).

Using the fact that $\int P(F,z) dF = \int P(E,z) dE$,
$ R(F,z) $ and $ P(E) $ are related by
\begin{eqnarray}
	R(F, z)
	&=& P(F|z) R(z) = R(E, z) \frac{dE}{dF} \\
	\label{eqn:pFz}
	&=& P(E) R(z) \frac{dE}{dF}.
\end{eqnarray}

$ R(z) $ is a rate per unit redshift, and contains the comoving volume element scaled by the star formation rate (SFR) at that given redshift:
\begin{equation}\label{eqn:R(z)}
	R(z) = \chi(z)^2 \frac{d\chi}{dz} \Phi(z)
\end{equation}
Since $\chi(z)$ is the comoving distance, the component $\chi(z)^2 \frac{d\chi}{dz}$ is equivalent to a volume element $\frac{dV}{dz}$.
$\Phi(z)$ is the rate of FRBs per comoving volume, which can then be generically modeled as 
\begin{equation}\label{eqn:Phi(z)}
	\Phi(z) = \frac{\Phi_0}{1+z}
	\left(\frac{\sfr(z)}{\sfr(0)} \right)^n,
\end{equation}
with $\Phi_0$
representing the rate of FRBs per comoving volume at $z=0$ and $E > E_\mathrm{pivot}$,
taking the units of bursts per proper time per comoving volume,
i.e., bursts yr$^{-1}$ Gpc$^{-1}$.

The star formation rate $\sfr(z)$ is given by \citet{madaudickinson2014} as
\begin{equation}\label{eqn:SFR(z)}
	\sfr(z) = 1.0025738 \frac{(1+z)^{2.7}}
	{1+\left(\frac{1+z}{2.9}\right)^{5.6}}.
\end{equation}
This means the distribution of FRB energy as a function of distance is proportional to the distribution of energies at a given volume element and scaled by the star formation rate SFR.

We can obtain $ \frac{dE}{dF} $ from Equation~\ref{eqn:E(F)}, which we can then combine with Equations~\ref{eqn:schechter_appendix} and~\ref{eqn:R(z)} to get
\begin{equation}\label{eqn:R(F,z)}
	R(F, z)
	= P(E) \chi(z)^2 \frac{d\chi}{dz} \Phi(z)
	\frac{4 \pi D^2_L(z)}{(1+z)^{2+\alpha}} \Delta \nu.
\end{equation}

Thus, the parameters in $\bm{\lambda_1}$ are $\alpha$, the spectral index; $n$, the power-law scaling with SFR; and $\gamma$ and $E_\mathrm{char}$, the power-law index and exponential cutoff energy which together characterize the Schechter energy function.

This aspect of the model was evaluated on an intrinsic grid in $F$ and $z$ space.
There were 200 $F$ points spaced logarithmically from $F = 1$--$5 \times 10^5$ Jy ms, encompassing the range of burst fluences from the Catalog~1 sample.
There were 100 points in $z$ space, spaced logarithmically from $z = 1 \times 10^{-3}$--1 and then linearly from $z = 1$--4.
This sampling was done in order to densely sample and reduce numerical error for the relatively low-$z$ region, where we would expect most FRBs to be.

\subsection{DM distribution}
\label{sec:P(DM|z)}

We now have the first component of the integrand needed to find $ R(F, \dm | \bm{\lambda}) $; the second component of the integrand is given by
\begin{equation}\label{eqn:P(dm|z)}
    P(\dm | z, \bm{\lambda_2}) = P(\dm = \dm_{\mathrm{EG}} + \overline{\dm}_{\mathrm{MW}} | z, \bm{\lambda_2})
\end{equation}
where the observed $\dm$ has an extragalactic contribution $\dm_{\mathrm{EG}}$ and an average Galactic contribution $\overline{\dm}_{\mathrm{MW}}$ consisting of contributions from the disk and halo of the Milky Way.
To obtain $\dm_{\mathrm{disk}}$,
the interstellar medium contribution to the $\dm$ in the Milky Way disk,
we consider the 
NE2001 model of electron density \citep{ne2001}
and the YMW2016 model of electron density \citep{ymw2016}
for the observed Catalog~1 bursts.
The former gives a median value of 
$\dm_{\mathrm{disk}} \approx 50$ pc cm$^{-3}$,
while the latter gives a median value of
$\dm_{\mathrm{disk}} \approx 45$ pc cm$^{-3}$.
The halo contribution is much more uncertain, with estimates ranging from $\approx$10--80 pc cm$^{-3}$
\citep[e.g.,][]{dolag2015, prochaskazheng2019, keatingpen2020}.
As done by \citet{chawla2022}, we start with the NE2001 model for estimating $\dm_{\mathrm{disk}}$,
and based on \citet{dolag2015}, adopt $\dm_{\mathrm{halo}}$ = 30 pc cm$^{-3}$.
Thus, $\overline{\dm}_{\mathrm{MW}}$ is originally fixed to be 80 pc cm$^{-3}$ in our model, where deviations from the assumption will be absorbed into contributions from the host galaxy of the FRB.
We test the sensitivity of our results to the adopted $\overline{\dm}_{\mathrm{MW}}$ in Section~\ref{sec:varyDML}.
Since $\overline{\dm}_{\mathrm{MW}}$ is fixed, obtaining $P(\dm_\mathrm{EG}|z, \bm{\lambda_2})$ will allow us to then obtain $ P(\dm|z, \bm{\lambda_2}) $. 

$\dm_{\mathrm{EG}}$ has contributions from the cosmological ionized gas distribution and host galaxy of the FRB, e.g., $\dm_{\mathrm{EG}} = \dm_{\mathrm{cosmic}} + \dm_{\mathrm{host}}$.
Using the mean density of ions $\bar{n}_e$ and relevant cosmological parameters, the expected cosmological contribution to the $\dm$ can be written as
\begin{equation}
    \label{eqn:dmcosmic}
    \left< \dm_{\mathrm{cosmic}} \right>(z)
    = \int_0^z
    \frac{c \bar{n}_e(z') dz'}
    {H_0 (1+z')^2 \sqrt{\Omega_m (1+z')^3 + \Omega_\Lambda}}.
\end{equation}
Then 
\begin{equation}
    \dm_{\mathrm{cosmic}} = \Delta_\dm \left< \dm_{\mathrm{cosmic}} \right>(z),
\end{equation}
where $\Delta_\dm$ is the deviation from the mean.
Thus, 
\begin{eqnarray}
    P(\dm_{\mathrm{cosmic}}|z)
    &=& P(\Delta_\dm) 
    \frac{d \Delta_\dm}{d \dm_{\mathrm{cosmic}}} \\
    &=& P(\Delta_\dm) 
    \frac{1}{\left< \dm_{\mathrm{cosmic}} \right>(z)}.
\end{eqnarray}

The expression $P(\Delta_\dm)$ is given by
\begin{equation}
    P(\Delta_\dm | z) =
    \kappa \Delta_\dm^{-B}
    \exp{\left[
    -\frac{(\Delta_\dm^{-A} - C_0)^2}{2 A^2 \sigma^2_\dm}
    \right]}
\end{equation}
with $A=3$, $B=3$, $\kappa$ the normalization factor, $C_0$ set by the requirement of $\langle \Delta_\dm \rangle = 1$, and $\sigma_\dm$ parameterized by feedback with $F z^{-0.5} = 0.32 z^{-0.5}$ \citep{james+2021_zDM}.
Note that these parameters are fixed in all the fits.

The host galaxy contribution can be generically modeled as a log-normal distribution
\begin{equation}
    P(\dm_{\mathrm{host}}') =
    \frac{1}{\dm_{\mathrm{host}}'}
    \frac{1}{\sigma_{\mathrm{host}} \sqrt{2 \pi}}
    \exp{\left[
    -\frac{(\log \dm_{\mathrm{host}}' - \mu_{\mathrm{host}})^2}
    {2 \sigma^2_{\mathrm{host}}}
    \right]},
\end{equation}
where $\mu_{\mathrm{host}}$ and $\sigma_{\mathrm{host}}$ are free parameters that make up the parameter vector $\bm{\lambda_2}$.
The median of this distribution is $\exp(\mu_{\mathrm{host}})$, and the variance is $[\exp(\sigma_{\mathrm{host}}^2)-1] \exp(2 \mu_{\mathrm{host}} + \sigma_{\mathrm{host}}^2)$.
Thus, $\bm{\lambda_2}$ contains information related to the host galaxy contribution to the extragalactic $\dm$.

The host contribution should also be corrected for redshift via
\begin{equation}
    \dm_{\mathrm{host}} = \frac{\dm_{\mathrm{host}}'}{1+z}.
\end{equation}
A more sophisticated model, using a dataset with more constraining power, would also take into consideration the evolution of the $\dm$ distribution with the cosmic star formation history \citep[e.g.,][]{luo2018lumin}.

Similarly to the cosmological $\dm$,
\begin{eqnarray}
    P(\dm_{\mathrm{host}}|z)
    &=& P(\dm_{\mathrm{host}}') 
    \frac{d \dm_{\mathrm{host}}'}{\dm_{\mathrm{host}}} \\
    &=& \frac{1}{\dm_{\mathrm{host}}'}
    \frac{1}{\sigma_{\mathrm{host}} \sqrt{2 \pi}}
    \exp{\left[
    -\frac{(\log \dm_{\mathrm{host}}' - \mu_{\mathrm{host}})^2}
    {2 \sigma^2_{\mathrm{host}}}
    \right]}
    (1+z).
\end{eqnarray}

With expressions for $P(\dm_{\mathrm{cosmic}}|z)$ and $P(\dm_{\mathrm{host}}|z)$, the expression for the extragalactic contribution to the $\dm$ becomes a convolution:
\begin{equation}
    P(\dm_{\mathrm{EG}} | z)
    = \int d\dm_{\mathrm{host}} 
    P(\dm_{\mathrm{host}}|z)
    P(\dm_{\mathrm{cosmic}} 
    = \dm_\mathrm{EG} - \dm_\mathrm{host}|z).
\end{equation}

However, our expression for $P(\dm_\mathrm{host}|z)$ is analytical, while our expression for $P(\dm_\mathrm{cosmic}|z)$ is numerical. Thus, a change of variables in the convolution is better for ease of evaluation:
\begin{equation}
    P(\dm_{\mathrm{EG}} | z)
    = \int d\dm_{\mathrm{cosmic}} 
    P(\dm_{\mathrm{host}}= \dm_\mathrm{EG} - \dm_\mathrm{cosmic}|z)
    P(\dm_{\mathrm{cosmic}}|z).
\end{equation}

Note that while \citet{chawla2022} modeled host galaxy contributions to the $\dm$ using the Catalog~1 sample, the results of the simulations were predictions with noise.
While it was the more complete way to model host galaxy contribution compared to this model,
this methodology needs a more precise analytical form.

By Equation~\ref{eqn:P(dm|z)}, we can obtain $ P(\dm|z) $.
Combining that with the earlier expression for $R(F,z)$ gives us the brightness distribution $R(F, \dm)$.

This aspect of the model was evaluated on an intrinsic grid in $z$ and $\dm$ space.
The points in $z$ space are the same as outlined in Section~\ref{sec:R(F,z)}, whereas there are 150 points in $\dm$ space linearly spaced from $\dm = $ 100--6500 pc cm$^{-3}$.
When evaluating on the $\dm_{\mathrm{EG}}$ grid,
$\overline{\dm}_{\mathrm{MW}}$ is subtracted from the $\dm$.

\section{Supplemental results for the ``rate interpretation'' of $\alpha$}
\label{sec:rate_alpha_supplementals}

\begin{figure*}[htbp]
    \centering
    \includegraphics[width=\textwidth]{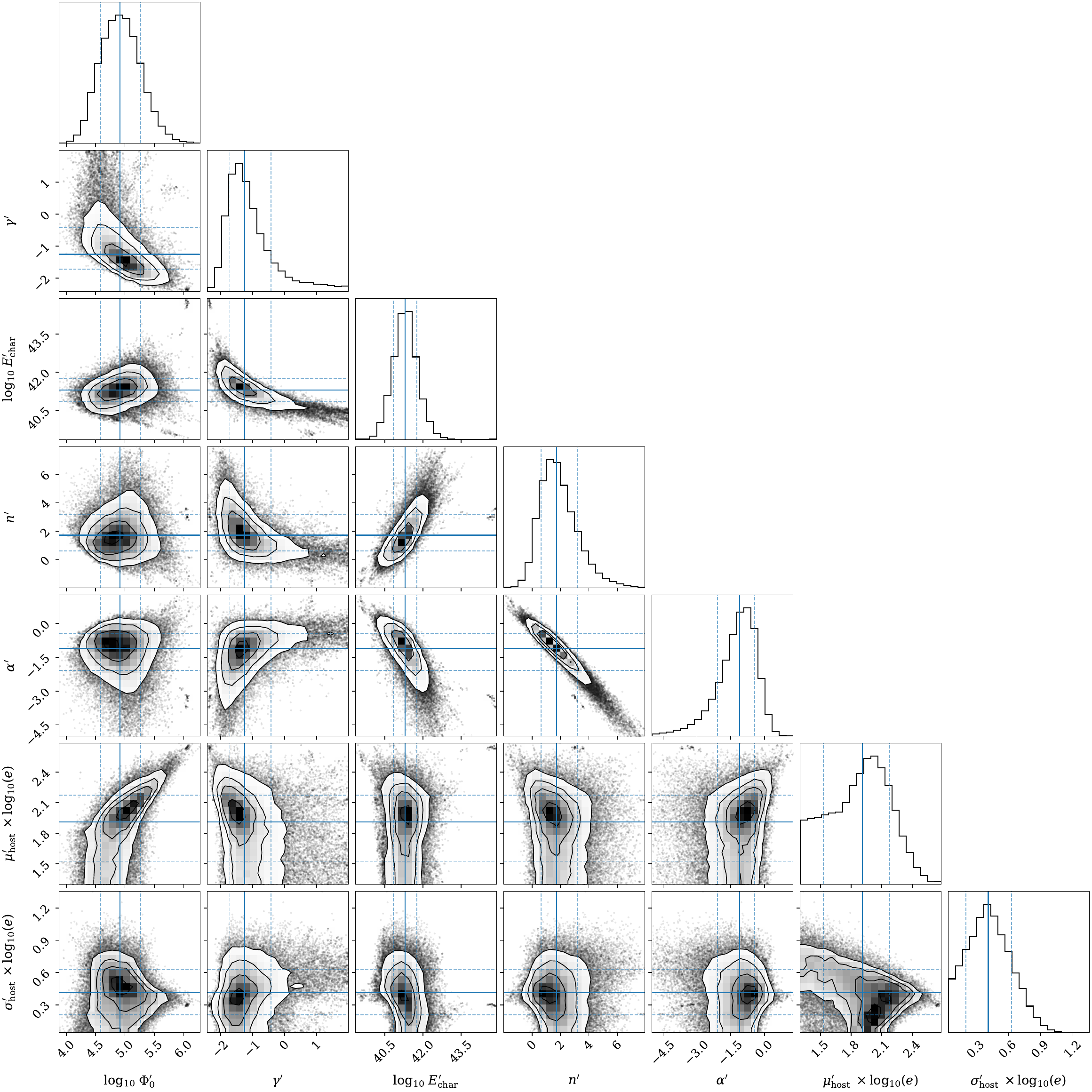}
    \caption{
        Corner plot of the results of the MCMC run under the ``rate interpretation'' of $\alpha$, thinned by a factor of 20 for visualization purposes.
        Overlaid solid blue lines denote the median of the posterior distributions, with the dashed blue lines enclosing the central 68\% of the samples.
    }
    \label{fig:mcmc_corner_alpharate}
\end{figure*}

This section contains the equivalent results of Figure~\ref{fig:mcmc_corner} and Table~\ref{tab:param_results},
except with the analysis done under the ``rate interpretation'' of $\alpha$.
They are presented in Figure~\ref{fig:mcmc_corner_alpharate} and Table~\ref{tab:param_results_alpharate}.
The logic in following few paragraphs was originally presented in \citet{macquartekers2018_fluence}, and are presented again here to emphasize how the equations change between the ``rate interpretation'' of $\alpha$ and the ``true spectral index interpretation''.

\begin{deluxetable}{ccc}
    \tablecaption{Table of results of the parameters fit in $R(F, \dm)$ assuming the ``rate'' interpretation of $\alpha$.
    Best-fit results quote the median values of the posterior distributions, with the error bars containing the central 68\% of the samples. Where illustrative, fits are paired with corresponding physical values.}
    \label{tab:param_results_alpharate}
    \tablehead{\colhead{Parameter} & \colhead{Uniform prior range} & \colhead{Best-fit result}} 
    
    \startdata
$\log_{10} \; \Phi_0\sp{\prime} \: ^a$ & [$-0.96$, $6.43$] & $4.83^{+0.35}_{-0.33}$ \\
$\Phi_0\sp{\prime}$ & $\ldots$ & $6.8^{+8.3}_{-3.6} \times 10^{4}$ Gpc$^{-3}$ yr$^{-1}$ \\[0.5em]
$\gamma\sp{\prime}$ & [$-2.50$, $2.00$] & $-1.3^{+0.8}_{-0.5}$ \\[0.5em]
$\log_{10} \; E_{\mathrm{char}}\sp{\prime} \: ^a$ & [$38.00$, $49.00$] & $41.29^{+0.47}_{-0.46}$ \\
$E_{\mathrm{char}}\sp{\prime}$ & $\ldots$ & $1.95^{+3.75}_{-1.28} \times 10^{41}$ erg \\[0.5em]
$n\sp{\prime}$ & [$-2.00$, $8.00$] & $1.72^{+1.48}_{-1.10}$ \\[0.5em]
$\alpha\sp{\prime}$ & [$-5.00$, $5.00$] & $-1.10^{+0.67}_{-0.99}$ \\[0.5em]
$\mu_{\mathrm{host}}\sp{\prime} \; \times \log_{10}(e) \: ^a$ & [$1.30$, $2.70$] & $1.91^{+0.27}_{-0.38}$ \\
$\mathrm{Median}\sp{\prime} \; P(\mathrm{DM}\sp{\prime})$ & $\ldots$ & $81^{+69}_{-48} $ pc cm$^{-3}$ \\[0.5em]
$\sigma_{\mathrm{host}}\sp{\prime} \; \times \log_{10}(e) \: ^a$ & [$0.04$, $1.74$] & $0.41^{+0.22}_{-0.21}$ \\
$\mathrm{Std. dev.}\sp{\prime} \; P(\mathrm{DM}\sp{\prime})$ & $\ldots$ & $174^{+347}_{-130}$ pc cm$^{-3}$ \\
    \enddata
    \tablenotetext{a}{These parameters were fit in natural $\log$ space but are presented in $\log_{10}$ space for ease of interpretation.}

\end{deluxetable}

The flux density in the frame of the observer is related to the luminosity in the frame of the emitting source by the following relation:
\begin{equation}
    S_\nu = (1+z) \frac{L_{(1+z) \nu}}{L_\nu} \frac{L_\nu}{4 \pi D_L^2},
\end{equation}
where $D_L$ is the luminosity distance,
and the ratio $L_{(1+z) \nu} / L_\nu$ represents the $k$-correction, necessary because the radiation was emitted in a different band than it was observed in.
Since $S_\nu \sim \nu^\alpha$, this $k$-correction factor is $(1+z)^\alpha$.

Observed fluence and emitted luminosity are related via the following relation \citep{maraninemiroff1996}:
\begin{equation}
    F_\nu = \Delta t_e (1+z)^2 \frac{L_{(1+z) \nu}}{L_\nu} \frac{L_\nu}{4 \pi D_L^2},
\end{equation}
where $\Delta t_e$ represents the burst duration in the emission frame.
Given that $E_\nu = \int L_\nu(t) dt$, and we are only considering $E$ and $F$ within a frequency bandwidth $\Delta \nu$, we can rewrite the above relation as
\begin{equation}
    F = (1+z)^2 (1+z)^\alpha \frac{E}{4 \pi D_L^2 \Delta \nu}.
\end{equation}
With some rearranging, this equation is equivalent to Equation~\ref{eqn:E(F)}.
This factor of $\alpha$ is explicitly due to the $k$-correction from the ``true spectral index interpretation'' of $\alpha$, i.e., $F \sim \nu^\alpha$.

Under the ``rate interpretation'' of $\alpha$, the $k$-correction instead comes into play by directly modifying the rate.
As mentioned in Section~\ref{sec:rate_interp_alpha}, Equation~\ref{eqn:Phi(z)} is modified such that it becomes 
\begin{equation}
    \Phi(z) = \frac{\Phi_0}{1+z}
	\left(\frac{\sfr(z)}{\sfr(0)} \right)^n (1+z)^{\alpha}.
\end{equation}
Additionally, Equation~\ref{eqn:E(F)} is now
\begin{equation}
    E = \frac{4 \pi D^2_L(z)}{(1+z)^{2}} \Delta \nu F.
\end{equation}

It is worth noting that although the parameter values of $n$ and $\alpha$ changed under this interpretation, the constraints on both values were not very powerful to begin with.
We present these results for completeness regarding systematic modeling uncertainties.

\section{Testing a FRB evolution model delayed with respect to SFR}
\label{sec:time_delay_sfrmodel}

In our fiducial $R(F, \dm)$ model, the evolution of the FRB population smoothly scales with the SFR as a power-law parameterized by $n$:
\begin{equation}
    \Phi(z) = \frac{\Phi_0}{1+z} \left( \frac{\sfr(z)}{\sfr(0)} \right)^n.
\end{equation}

\begin{figure*}[htbp]
    \centering
    \includegraphics[width=\textwidth]{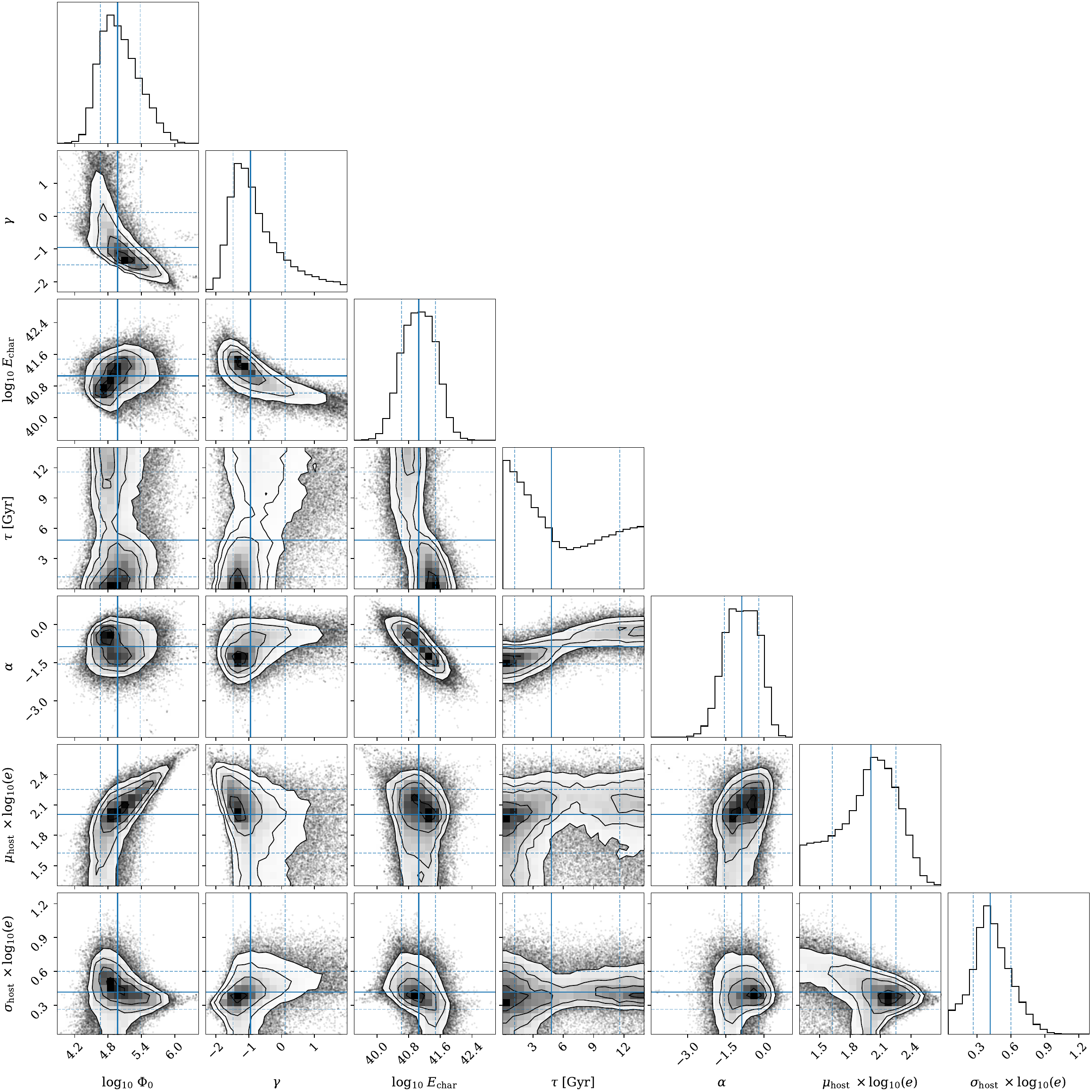}
    \caption{
        Corner plot of the results of the MCMC run using a time-delay toy model, thinned by a factor of 20 for visualization purposes.
        Overlaid solid blue lines denote the median of the posterior distributions, with the dashed blue lines enclosing the central 68\% of the samples.
    }
    \label{fig:mcmc_corner_timedelaysfr}
\end{figure*}

Despite the poor constraints on $n$, we decide to explore whether our data can constrain FRB population in a toy model where the rate of FRBs strictly lags behind the SFR by a delay parameter $\tau$~Gyr.
Such a model is represented by
\begin{equation}
    \Phi(z) = \frac{\Phi_0}{1+z} \frac{\sfr( z_a(t_\mathrm{MW}(z) - \tau) )}{\sfr(0)},
\end{equation}
where $z_a$ is the function that obtains the redshift at a given lookback time.
Of course, it is not physical to expect that for each star that forms, an FRB occurs exactly $\tau$~Gyr later.
This toy model is intended to be for illustrative purposes only, as its key purpose is to investigate whether our data have any constraining power for a more complex FRB population evolution model.
Such a more realistic model would include a distribution of delay times, as in \citet{zhangzhang22}.
If this toy model shows no constraining power, then a more complex model would almost certainly have no constraining power as well.
Thus, we run this toy model first with the same procedure as in Sections~\ref{sec:methodology} and \ref{sec:results}, with uniform priors on $\tau$ from 0 -- 14~Gyr.
The results of the MCMC are plotted in Figure~\ref{fig:mcmc_corner_timedelaysfr}.

We immediately see that $\tau$ is even more poorly constrained than $n$; in other words, with this toy model, our data still hold no constraining power for the FRB evolution with SFR.
The posterior distribution for $\tau$ appears bimodal, with peaks at both negligible ($\sim$0~Gyr) time delay and significant ($\sim$12~Gyr) time delay.
Of our $10^6$ chains, $\approx$25.8\% have $\tau < 2$~Gyr and $\approx$13.3\% have $12$~Gyr~$< 14$~Gyr (the rest of the parameter results vary across chains).
Although it is tempting to ascribe significance to this bimodality, we caution against over-interpretation for a number of reasons.
First, the middle of the posterior distribution does not dip down to negligible probabilities, mitigating the bimodal appearance of this distribution.
Second, and more importantly, this is an unphysical toy model whose primary purpose was to explore whether the Catalog~1 data could constrain a model where FRBs as a population are delayed with respect to the SFH of the universe.
We conclude that the Catalog~1 data cannot constrain such a model.

\bibliographystyle{aasjournal}
\bibliography{refs}

\end{document}